\documentclass[aps,twocolumn,showpacs]{revtex4}
\usepackage{epsf}
\usepackage{bm}

\begin{document}

\title{van der Waals coupling in atomically doped carbon nanotubes}

\author{I.V.~Bondarev\footnote{Corresponding author. E-mail address:
{\sf bondarev@tut.by}}}
\affiliation{Institute for Nuclear Problems, Belarusian
State University, Bobruiskaya Str.11, 220050 Minsk, BELARUS}
\author{Ph.~Lambin}
\affiliation{Facult\'{e}s Universitaires Notre-Dame de la Paix, 61
rue de Bruxelles, 5000 Namur, BELGIUM}

\begin{abstract}
We have investigated atom-nanotube van der Waals (vdW) coupling in
atomically doped carbon nanotubes (CNs).~Our approach is based on
the perturbation theory for degenerated atomic levels, thus
accounting for both weak and strong atom-vacuum-field
coupling.~The vdW energy is described by an integral equation
represented in terms of the local photonic density of states
(DOS).~By solving it numerically, we demonstrate the
inapplicability of standard weak-coupling-based vdW interaction
models in a close vicinity of the CN surface where the local
photonic DOS effectively increases, giving rise to an atom-field
coupling enhancement.~An inside encapsulation of atoms into the CN
has been shown to be energetically more favorable than their
outside adsorption by the CN surface.~If the atom is fixed outside
the CN, the modulus of the vdW energy increases with the CN radius
provided that the weak atom-field coupling regime is realized
(i.e.,~far enough from the CN).~For inside atomic position, the
modulus of the vdW energy decreases with the CN radius,
representing a general effect of the effective interaction area
reduction with lowering the CN curvature.
\end{abstract}


\maketitle

\section{Introduction}

Carbon nanotubes (CNs) are graphene sheets rolled-up into
cylinders of approximately one nanometer in diameter.~Extensive
work carried out worldwide in recent years has revealed the
intriguing physical properties of these novel molecular scale
wires~\cite{Dresselhaus,Dai}.~Nanotubes have been shown to be
useful for miniaturized electronic, mechanical, electromechanical,
chemical and scanning probe devices and materials for macroscopic
composites~\cite{Baughman}.~Important is that their intrinsic
properties may be substantially modified in a controllable way by
doping with extrinsic impurity atoms, molecules and
compounds~\cite{Duclaux}.~Recent successful experiments on
encapsulation of single atoms into single-wall CNs~\cite{Jeong}
and their intercalation into single-wall CN
bundles~\cite{Duclaux,Shimoda}, along with numerous studies of
monoatomic gas absorbtion by the CN bundles (see~\cite{Calbi} for
a review), stimulate an in-depth analysis of the atom-CN van der
Waals (vdW) interactions.~The problem is both of fundamental and
of applied interest, because it sheds light on the peculiarities
of the atom-electromagnetic-field interactions in low-dimensional
dispersive and absorbing surroundings, on the one hand, and, on
the other, the atom-nanotube vdW interaction is the most important
one determining the filling factor in encapsulating single atoms
into CNs.

The vdW interaction is the result of electrodynamic coupling
between the polarization states of the interacting entities that
results from the vacuum fluctuations of the electromagnetic
field.~A theoretical analysis recently
done~\cite{Bondarev02,Bondarev04PLA,Bondarev04PRB} of spontaneous
decay dynamics of excited atomic states near CNs has brought out
fascinating peculiarities of the vacuum-field interactions in
atomically doped CNs.~In particular, close to the nanotube, the
relative density of photonic states (DOS) effectively increases
due to the presence of additional surface photonic states coupled
with CN electronic quasiparticle excitations.~This causes an
atom-vacuum-field coupling constant, which is proportional to the
photonic DOS, to be very sensitive to an atom-CN-surface
distance.~If the atom is close enough to the CN surface and the
atomic transition frequency is in the vicinity of a resonance of
the photonic DOS, the system shows strong atom-vacuum-field
coupling giving rise to rearrangement ("dressing") of atomic
levels by the vacuum-field interaction.~If the atom moves away
from the CN surface, the atom-field coupling strength decreases,
smoothly approaching the weak coupling regime at large
atom-surface distances since the role of surface photonic states
diminishes, causing the relative photonic DOS to decrease.~This
suggests strictly nonlinear atom-field coupling and a primary role
of the distance-dependent (local) photonic DOS in the vicinity of
the CN, so that (weak-coupling) vdW interaction models based upon
standard vacuum quantum electrodynamics (QED) as well as those
using the linear response theory
(see~\cite{Buhmann04JOB,Buhmann04pre} for a review) are, in
general, inapplicable for an atom in a close vicinity of a carbon
nanotube.

To give this issue a proper quantum electrodynamic consideration
and thereby to advance in physical understanding the vdW
interactions deeper than simple analyzing empirical model
potentials~\cite{Girifalco}, we develop a~universal quantum
approach to the vdW energy calculation of a (two-level) atomic
system in the vicinity of an infinitely long single-wall CN.~The
approach is based upon the perturbation theory for degenerate
atomic levels (see, e.g.,~\cite{Davydov}), thereby allowing one to
account for both strong and weak atom-vacuum-field coupling
regimes.~We derive an integral equation for the vdW energy of the
two-level atomic system near the CN.~The equation is represented
in terms of the local photonic DOS and is valid for both strong
and weak atom-field coupling.~By solving it numerically, we
demonstrate the inapplicability of weak-coupling-based vdW
interaction models in a close vicinity of the nanotube surface.~We
also show that an inside encapsulation of doped atoms into the
nanotube is energetically more favorable than their outside
adsorption by the nanotube surface.

The rest of the paper is arranged as
follows.~Section~\ref{vdwenergy} presents a theoretical model we
use to derive a~universal integral equation for the ground-state
vdW energy of the two-level atomic system interacting with
a~vacuum electromagnetic field in the vicinity (inside or outside)
of the CN. In describing the atom--field interaction, we follow an
original line of Ref.~\cite{Welsch,Dung}, adopting their
electromagnetic field quantization scheme for a~particular case of
the field near an infinitely long achiral single-wall CN.~The
carbon nanotube is considered to be an infinitely thin
anisotropically conducting cylinder.~Its surface conductivity is
represented in terms of the $\pi$-electron dispersion law obtained
in the tight-binding approximation with allowance made for
azimuthal electron momentum quantization and axial electron
momentum relaxation~\cite{Slepyan}.~Only the axial conductivity is
taken into account and the azimuthal one, being strongly
suppressed by transverse depolarization
fields~\cite{Tasaki,Li,Jorio,Marinop}, is neglected.~In
Sec.~\ref{qualitatively}, the vdW energy equation derived in
Sec.~\ref{vdwenergy} is analyzed qualitatively to demonstrate its
validity and correctness in both weak and strong atom-field
coupling regimes.~We show that the equation obtained reproduces a
well-known perturbation theory result in the weak coupling regime
and reduces to the integral equation of a simplified form in the
strong coupling regime.~Section~\ref{numerically} presents and
discusses the results of the various numerical calculations of the
vdW energies of the two-level atom placed inside or outside
achiral CNs of different radii.~A summary and conclusions of the
work are given in Sec.~\ref{conclusion}.

\section{The ground-state van der Waals energy}\label{vdwenergy}

\subsection{The field quantization formalism}\label{quantization}

The quantum theory of the van der Waals interaction between an
atomic system and a carbon nanotube involves an electromagnetic
field quantization procedure in the presence of dispersing and
absorbing bodies.~Such~a procedure faces difficulties similar to
those in quantum optics of 3D Kramers-Kronig dielectric media
where the canonical quantization scheme commonly used does not
work since, because of absorption, corresponding operator Maxwell
equations become non-Hermitian~\cite{Vogel}.~As a consequence,
their solutions cannot be expanded in power orthogonal modes and
the concept of modes itself becomes more subtle.~We, therefore,
use a unified macroscopic QED approach developed in
Refs.~\cite{Dung,Welsch} and adapted for an atom near a CN in
Refs.~\cite{Bondarev04PLA,Bondarev04PRB}.~In this approach, the
Fourier-images of electric and magnetic fields are considered as
quantum mechanical observables of corresponding electric and
magnetic field operators.~The latter ones satisfy the
Fourier-domain operator Maxwell equations modified by the presence
of a so-called operator noise current density
$\underline{\hat{\mathbf{J}}\!}\,(\mathbf{r},\omega)$ written in
terms of a three-dimensional (3D) vector bosonic field operator
$\hat\mathbf{f}(\mathbf{r},\omega)$ and a medium dielectric tensor
$\bm\epsilon(\mathbf{r},\omega)$ (supposed to be diagonal) as
\begin{equation}
{\underline{\hat{J}\!}_{\,i}}(\mathbf{r},\omega)={\omega\over{2\pi}}
\sqrt{\hbar\,\mbox{Im}\epsilon_{ii}(\mathbf{r},\omega)}\,
\hat{f}_{i}(\mathbf{r},\omega),\hskip0.1cm i=1,2,3.
\label{current}
\end{equation}
This operator is responsible for correct commutation relations of
the electric and magnetic field operators in the presence of
medium-induced absorbtion.~In this formalism, the electric and
magnetic field operators are expressed in terms of a continuum set
of the 3D vector bosonic fields
$\hat{\mathbf{f}}(\mathbf{r},\omega)$ by means of the convolution
over~$\mathbf{r}$ of the current (\ref{current}) with the
classical electromagnetic field Green tensor of the system.~The
bosonic field operators
$\hat{\mathbf{f}}^{\dag}(\mathbf{r},\omega)$ and
$\hat{\mathbf{f}}(\mathbf{r},\omega)$ create and annihilate
single-quantum electromagnetic medium excitations.~They are
defined by their commutation relations
\begin{eqnarray}
\mbox{[}\,\hat{f}_{i}(\mathbf{r},\omega),
\hat{f}^{\dag}_{j}(\mathbf{r}^{\prime},\omega^{\prime})\,\mbox{]}&=&
\delta_{ij}\delta(\mathbf{r}-\mathbf{r}^{\prime})\delta(\omega-\omega^{\prime})\,,
\label{commut}\\
\mbox{[}\,\hat{f}_{i}(\mathbf{r},\omega),
\hat{f}_{j}(\mathbf{r}^{\prime},\omega^{\prime})\,\mbox{]}&=&
\mbox{[}\,\hat{f}^{\dag}_{i}(\mathbf{r},\omega),
\hat{f}^{\dag}_{j}(\mathbf{r}^{\prime},\omega^{\prime})\,\mbox{]}\,=\,0\nonumber
\end{eqnarray}
and play the role of the fundamental dynamical variables in terms
of which the Hamiltonian of the composed system "electromagnetic
field + dissipative medium" is written in a standard secondly
quantized form as
\begin{equation}
\hat{H}=\int_{0}^{\infty}\!\!\!d\omega\,\hbar\omega\!\int\!d\mathbf{r}\,
\hat{\mathbf{f}}^{\dag}(\mathbf{r},\omega)\!
\cdot\!\hat{\mathbf{f}}(\mathbf{r},\omega)\,. \label{H_field}
\end{equation}

\begin{figure}[t]
\epsfxsize=8.65cm\centering{\epsfbox{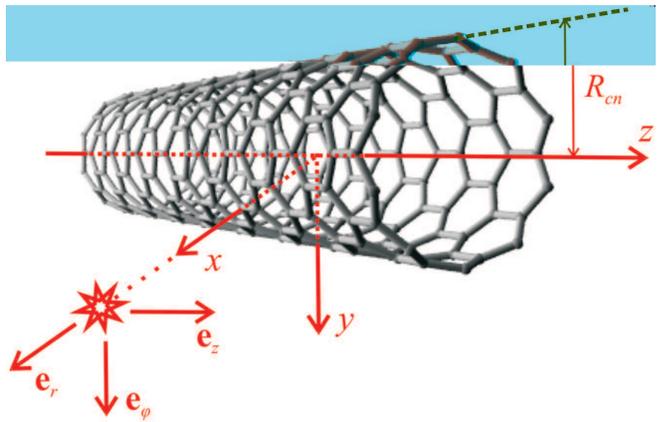}} \caption{(Color
online) The geometry of the problem.} \label{fig1}
\end{figure}

Consider a neutral atomic system with its centre-of-mass
positioned at the point $\mathbf{r}_{A}$ near an infinitely long
single-wall CN. Since the problem has a cylindric symmetry, it is
convenient to assign the orthonormal cylindric basis
$\{\mathbf{e}_{r},\mathbf{e}_{\varphi},\mathbf{e}_{z}\}$ in such a
way that $\mathbf{e}_{z}$ is directed along the nanotube axis and
$\mathbf{r}_{A}=r_{A}\mathbf{e}_{r}=\{r_{A},0,0\}$~(see
Fig.~\ref{fig1}).~For carbon nanotubes, their strong transverse
depolarization along with transverse electron momentum
quantization allow one to neglect the azimuthal current and radial
polarizability~\cite{Slepyan,Tasaki,Jorio,Li,Marinop}, in which
case the dielectric tensor components $\epsilon_{rr}$ and
$\epsilon_{\varphi\varphi}$ are identically equal to unit.~The
component $\epsilon_{zz}$ is caused by the CN longitudinal
polarizability and is responsible for the axial surface current
parallel to the $\textbf{e}_{z}$ vector.~This current may be
represented in terms of the 1D bosonic field operators by analogy
with Eq.~(\ref{current}).~Indeed, taking into account the
dimensionality conservation in passing from bulk to a~monolayer in
Eq.~(\ref{H_field}) and using a simple Drude
relation~\cite{Tasaki}
\begin{equation}
\sigma_{zz}(\mathbf{R},\omega)=-i\omega{\epsilon_{zz}(\mathbf{R},\omega)-
1\over{4\pi S\rho_{T}}}\,, \label{sigmaCN}
\end{equation}
where $\mathbf{R}=\!\{R_{cn},\phi,Z\}$~is the radius-vector of an
arbitrary point of the CN surface, $R_{cn}$ is the radius of the
CN, $\sigma_{zz}(\mathbf{R},\omega)$ is the CN surface axial
conductivity per unit length, $S$ is the area of a single
nanotube, $\rho_{T}$ is the tubule density in a bundle, one
immediately has from Eq.~(\ref{current})
\begin{equation}
\underline{\hat{\mathbf{J}}\!}\,(\mathbf{R},\omega)\!=
\!\sqrt{\hbar\omega\mbox{Re}\sigma_{zz}(\mathbf{R},\omega)\over{\pi}}\,
\hat{f}(\mathbf{R},\omega)\textbf{e}_{z} \label{currentCN}
\end{equation}
with $\hat{f}(\mathbf{R},\omega)$ being the 1D bosonic field
operator defined on the CN surface.~The total nonrelativistic
Hamiltonian of an atomic system interacting with a CN modified
quantized vacuum electromagnetic field can then be represented in
the form
\begin{eqnarray}
\hat{H}&=&\int_{0}^{\infty}\!\!\!\!\!d\omega\,\hbar\omega\!\int\!d\mathbf{R}
\,\hat{f}^{\dag}(\mathbf{R},\omega)\hat{f}(\mathbf{R},\omega)\nonumber\\
&+&\sum_{i}{1\over{2m_{i}}}\left[\hat{\mathbf{p}}_{i}-
{q_{i}\over{c}}\hat{\mathbf{A}}(\hat{\mathbf{r}}_{i})\right]^{2}\nonumber\\
&+&{1\over{2}}\int\!d\mathbf{r}\,\hat{\rho}_{A}(\mathbf{r})\hat{\varphi}_{A}(\mathbf{r})
+\int\!d\mathbf{r}\,\hat{\rho}_{A}(\mathbf{r})\hat{\varphi}(\mathbf{r})\,,
\label{Htotini}
\end{eqnarray}
where $m_{i}$, $q_{i}$, $\hat{\mathbf{r}}_{i}$ and
$\hat{\mathbf{p}}_{i}$ are, respectively, the masses, charges,
coordinates (relative to $\mathbf{r}_{A}$) and momenta of the
particles constituting the atomic subsystem.~The first term of the
Hamiltonian describes the \emph{medium-assisted} vacuum
electromagnetic field in the presence of the CN.~The second and
the third terms of the Hamiltonian represent the kinetic energy of
the charged particles and their mutual Coulomb interaction,
respectively, with
\begin{equation}
\hat{\varphi}_{A}(\mathbf{r})=\int\!d\mathbf{r}^{\prime}
{\hat{\rho}_{A}(\mathbf{r}^{\prime})\over{|\mathbf{r}-\mathbf{r}^{\prime}|}}
\label{phia}
\end{equation}
being the scalar potential and
\begin{equation}
\hat{\rho}_{A}(\mathbf{r})=\sum_{i}q_{i}\delta(\mathbf{r}-\hat{\mathbf{r}}_{i})
\label{rhoa}
\end{equation}
being the charged density of the atomic subsystem.~The last term
accounts for the Coulomb interaction of the particles with the CN.
The vector potential $\hat{\mathbf{A}}$ and the scalar potential
$\hat{\varphi}$ of the CN-modified electromagnetic field are given
for an arbitrary $\mathbf{r}=\{r,\varphi,z\}$ in the
Schr\"{o}dinger picture by
\begin{equation}
\hat{\mathbf{A}}(\mathbf{r})=\int_{0}^{\infty}\!\!\!\!\!d\omega{c\over{i\omega}}\,
\underline{\hat{\mathbf{E}}}^{\perp}(\mathbf{r},\omega)+\mbox{h.c.},
\label{vecpot}
\end{equation}
\begin{equation}
-\bm{\nabla}\hat{\varphi}(\mathbf{r})=\int_{0}^{\infty}\!\!\!\!\!d\omega\,
\underline{\hat{\mathbf{E}}}^{\parallel}(\mathbf{r},\omega)+\mbox{h.c.},
\label{scalpot}
\end{equation}
where
\begin{equation}
\underline{\hat{\mathbf{E}}}^{\perp(\parallel)}(\mathbf{r},\omega)=
\int\!d\mathbf{r}^{\prime}\,\bm{\delta}^{\perp(\parallel)}
(\mathbf{r}-\mathbf{r}^{\prime})\cdot\underline{\hat{\mathbf{E}}}
(\mathbf{r}^{\prime},\omega)
\label{elperpar}
\end{equation}
is the transverse (longitudinal) electric field operator with
\begin{equation}
\delta^{\parallel}_{\alpha\beta}(\mathbf{r})=
-\nabla_{\alpha}\nabla_{\beta}{1\over{4\pi r}}
\label{deltapar}
\end{equation}
and
\begin{equation}
\delta^{\perp}_{\alpha\beta}(\mathbf{r})=\delta_{\alpha\beta}\,
\delta(\mathbf{r})-\delta^{\parallel}_{\alpha\beta}(\mathbf{r})
\label{deltaper}
\end{equation}
being the longitudinal and transverse dyadic $\delta$-functions,
respectively, and $\underline{\hat{\mathbf{E}}}$ representing the
total electric field operator which satisfies the following set of
Fourier-domain Maxwell equations
\begin{equation}
\bm{\nabla}\times\underline{\hat{\mathbf{E}}}(\mathbf{r},\omega)=
ik\,\underline{\hat{\mathbf{H}}}(\mathbf{r},\omega),
\label{MaxwelE}
\end{equation}
\begin{equation}
\bm{\nabla}\times\underline{\hat{\mathbf{H}}}(\mathbf{r},\omega)=
-ik\,\underline{\hat{\mathbf{E}}}(\mathbf{r},\omega)+
{4\pi\over{c}}\underline{\hat{\mathbf{I}}}(\mathbf{r},\omega).
\label{MaxwelH}
\end{equation}
Here, $\underline{\hat{\mathbf{H}}}$ stands for the magnetic field
operator, $k=\omega/c$, and
\begin{eqnarray}
\underline{\hat{\mathbf{I}}}(\mathbf{r},\omega)&=&\int\!\!d\mathbf{R}\,
\delta(\mathbf{r}-\mathbf{R})\,\underline{\hat{\mathbf{J}}\!}\,(\mathbf{R},\omega)\nonumber\\
&=&2\underline{\hat{\mathbf{J}}\!}\,(R_{cn},\varphi,z,\omega)\,\delta(r-R_{cn})\,,
\label{Irw}
\end{eqnarray}
is the exterior operator current density [with
$\underline{\hat{\mathbf{J}}}(\mathbf{R},\omega)$ defined by
Eq.~(\ref{currentCN})] associated with the presence of the CN.

From Eqs.~(\ref{MaxwelE})-(\ref{Irw}) in view of
Eq.~(\ref{currentCN}), it follows that
\begin{equation}
\underline{\hat{\mathbf{E}}}(\mathbf{r},\omega)=
i{4\pi\over{c}}\,k\!\int\!\!d\mathbf{R}\,\mathbf{G}(\mathbf{r},\mathbf{R},\omega)
\!\cdot\!\underline{\hat{\mathbf{J}}\!}\,(\mathbf{R},\omega)
\label{Erw}
\end{equation}
[and $\underline{\hat{\mathbf{H}}}=(ik)^{-1}
\bm{\nabla}\times\underline{\hat{\mathbf{E}}}$ accordingly], where
$\mathbf{G}$ is the Green tensor of the classical electromagnetic
field in the vicinity of the CN.~The set of
Eqs.~(\ref{Htotini})-(\ref{Erw}) forms a closed electromagnetic
field quantization formalism in the presence of dispersing and
absorbing media which meets all the basic requirements of
a~standard quantum electrodynamics~\cite{Welsch}.~All information
about medium (a~CN in our case) is contained in the Green tensor
$\mathbf{G}$ via the CN surface axial conductivity in
Eq.~(\ref{currentCN}).~The Green tensor components satisfy the
equation
\begin{equation}
\sum_{\alpha=r,\varphi,z}\!\!\!
\left(\bm{\nabla}\!\times\bm{\nabla}\!\times-\,k^{2}\right)_{\!z\alpha}
G_{\alpha z}(\mathbf{r},\mathbf{R},\omega)=
\delta(\mathbf{r}-\mathbf{R}), \label{GreenequCN}
\end{equation}
together with the radiation conditions at infinity and the
boundary conditions on the CN surface. This tensor was derived and
analysed in Ref.~\cite{Bondarev04PRB}.

Assuming further that the atomic subsystem is sufficiently
localized in space, so that the long-wavelength approximation
applies, one can expand the field operators
$\hat{\mathbf{A}}(\mathbf{r})$ and $\hat{\varphi}(\mathbf{r})$ in
the Hamiltonian (\ref{Htotini}) around the atomic center of mass
position $\mathbf{r}_{A}$ and only keep the leading non-vanishing
terms of the expansions.~Then, under the condition of the Coulomb
gauge $[\hat{\mathbf{p}}_{i},\hat{\mathbf{A}}]=0$, one arrives at
the approximate total Hamiltonian of the form
\begin{equation}
\hat{H}=\hat{H}_{F}+\hat{H}_{A}+\hat{H}^{(1)}_{AF}+\hat{H}^{(2)}_{AF}\,,
\label{Htot}
\end{equation}
where
\begin{equation}
\hat{H}_{F}=\int_{0}^{\infty}\!\!\!\!\!d\omega\,\hbar\omega\!\int\!d\mathbf{R}
\,\hat{f}^{\dag}(\mathbf{R},\omega)\hat{f}(\mathbf{R},\omega)\,,
\label{Hf}
\end{equation}
\begin{equation}
\hat{H}_{A}=\sum_{i}{\hat{\mathbf{p}}^{2}\over{2m_{i}}}+
{1\over{2}}\sum_{i,j}{q_{i}q_{j}\over{|\mathbf{r}_{i}-\mathbf{r}_{j}|}}\,,
\label{Ha}
\end{equation}
\begin{equation}
\hat{H}^{(1)}_{AF}=-\sum_{i}{q_{i}\over{m_{i}c}}\,\hat{\mathbf{p}}_{i}
\!\cdot\!\hat{\mathbf{A}}(\mathbf{r}_{A})+\hat{\mathbf{d}}\!\cdot\!
\bm{\nabla}\hat{\varphi}(\mathbf{r}_{A})\,, \label{Haf1}
\end{equation}
\begin{equation}
\hat{H}^{(2)}_{AF}=\sum_{i}{q^{2}_{i}\over{2m_{i}c^{2}}}\,
\hat{\mathbf{A}}^{\!2}(\mathbf{r}_{A})
\label{Haf2}
\end{equation}
are, respectively, the Hamiltonian of the vacuum electromagnetic
field modified by the presence of the CN, the Hamiltonian of the
atomic subsystem, and the electric dipole approximation for the
Hamiltonian of the atom-field interaction (separated into two
contributions according to their role in the vdW energy -- see
below) with $\hat{\mathbf{d}}=\sum_{i}q_{i}\hat{\mathbf{r}}_{i}$
being the electric dipole moment operator of the atomic subsystem.

\subsection{The two-level approximation}\label{two-level}

Starting from Eqs.~(\ref{Htot})-(\ref{Haf2}) and using
Eqs.~(\ref{vecpot})-(\ref{deltaper}), (\ref{Erw}) and
(\ref{currentCN}), one obtains the following two-level
approximation for the total Hamiltonian of the system (see
Appendix~\ref{appA})
\begin{equation}
\hat{\cal{H}}=\hat{\cal{H}}_{0}+\hat{\cal{H}}_{int}\,,
\label{Htwolev}
\end{equation}
where the first term stands for the unperturbed Hamiltonian given
by
\begin{equation}
\hat{\cal{H}}_{0}=\hat{H}_{F}+\hat{\cal{H}}_{A}
\label{H0}
\end{equation}
with $\hat{H}_{F}$ being the Hamiltonian (\ref{Hf}) of the CN
modified field and
\begin{equation}
\hat{\cal{H}}_{A}={\hbar\tilde{\omega}_{A}\over{2}}\,\hat{\sigma}_{z}
\label{Harenorm}
\end{equation}
representing the 'effective' unperturbed two-level atomic
subsystem, and the second term stands for their interaction given
by
\begin{eqnarray}
\hat{\cal{H}}_{int}&=&\!\int_{0}^{\infty}\!\!\!\!\!d\omega\!\int\!d\mathbf{R}\;[\,
\mbox{g}^{(+)}(\mathbf{r}_{A},\mathbf{R},\omega)\,\hat{\sigma}^{\dag}\nonumber\\
&-&\mbox{g}^{(-)}(\mathbf{r}_{A},\mathbf{R},\omega)\,\hat{\sigma}\,]\,
\hat{f}(\mathbf{R},\omega)+\mbox{h.c.}. \label{Hint}
\end{eqnarray}
Here, the Pauli operators
\begin{eqnarray}
\hat{\sigma}=|l\rangle\langle u|\,,\hskip0.2cm
\hat{\sigma}^{\dag}=|u\rangle\langle l|\,,\hskip0.2cm
\hat{\sigma}_{z}&\!=\!&|u\rangle\langle u|-|l\rangle\langle l|\,,\label{Pauli}\\
|u\rangle\langle u|+|l\rangle\langle l|&\!=\!&\hat{I}\,,\nonumber
\end{eqnarray}
describe electric dipole transitions between the two ato\-mic
states, upper $|u\rangle$ and lower $|l\rangle$, separated by the
transition frequency $\omega_{A}$.~This 'bare' transition
frequency is modified by the interaction (\ref{Haf2}) which, being
independent of the atomic dipole moment, \emph{does not}
contribute to mixing the $|u\rangle$ and $|l\rangle$ states,
giving rise, however, to the new \emph{renormalized} transition
frequency
\begin{equation}
\tilde{\omega}_{A}=\omega_{A}\!\left[1-{2\over{(\hbar\omega_{A})^{2}}}
\int_{0}^{\infty}\!\!\!\!\!d\omega\!\int\!d\mathbf{R}\,
|\mbox{g}^{\perp}(\mathbf{r}_{A},\mathbf{R},\omega)|^{2}\right]
\label{omegarenorm}
\end{equation}
in the 'effective' atomic Hamiltonian~(\ref{Harenorm}).~On the
contrary, the interaction (\ref{Haf1}), being dipole moment
dependent, mixes the $|u\rangle$ and $|l\rangle$ states, yielding
the perturbation Hamiltonian~(\ref{Hint}) with the (dipole)
interaction matrix elements of the form
\begin{equation}
\mbox{g}^{(\pm)}(\mathbf{r}_{A},\mathbf{R},\omega)=
\mbox{g}^{\perp}(\mathbf{r}_{A},\mathbf{R},\omega)\pm
{\omega\over{\omega_{A}}}\,\mbox{g}^{\parallel}(\mathbf{r}_{A},\mathbf{R},\omega)\,,
\label{gpm}
\end{equation}
where
\begin{eqnarray}
\mbox{g}^{\perp(\parallel)}(\mathbf{r}_{A},\mathbf{R},\omega)=\;\;&&\nonumber\\
-i{4\omega_{A}\over{c^{2}}}\,d_{z}\sqrt{\pi\hbar\omega\,\mbox{Re}\,
\sigma_{zz}(\mathbf{R},\omega)}\;^{\perp(\parallel)}&\!\!\!\!
G_{zz}\!\!\!\!&(\mathbf{r}_{A},\mathbf{R},\omega)\mbox{\hskip0.5cm}
\label{gperppar}
\end{eqnarray}
with
\begin{equation}
^{\perp(\parallel)}G_{zz}(\mathbf{r}_{A},\mathbf{R},\omega)=\int\!d\mathbf{r}\,
\delta^{\perp(\parallel)}_{zz}(\mathbf{r}_{A}-\mathbf{r})\,
G_{zz}(\mathbf{r},\mathbf{R},\omega) \label{Gzzperpar}
\end{equation}
being the field Green tensor $zz$-component transverse
(longitudinal) with respect to the first variable.~This is the
only Green tensor component we have to take account of.~All the
other components can be safely be neglected, because our model
neglects the azimuthal current and radial polarizability of the CN
from the very outset.

The matrix elements (\ref{gpm})~and~(\ref{gperppar}) have the
following properties (see Appendix~\ref{appB})
\begin{equation}
\int\!d\mathbf{R}\,|\mbox{g}^{\perp(\parallel)}(\mathbf{r}_{A},\mathbf{R},\omega)|^{2}\!=
{(\hbar\omega_{A})^{2}\over{2\pi\omega^{2}}}\,\Gamma_{0}(\omega)\,
\xi^{\perp(\parallel)}(\mathbf{r}_{A},\omega)
\label{DOS}
\end{equation}
and
\begin{eqnarray}
\int\!d\mathbf{R}\,|\mbox{g}^{(\pm)}(\mathbf{r}_{A},\mathbf{R},\omega)|^{2}\!=
{(\hbar\omega_{A})^{2}\over{2\pi\omega^{2}}}\,\Gamma_{0}(\omega)\nonumber\\
\times\left[\xi^{\perp}(\mathbf{r}_{A},\omega)+
\!\left({\omega\over{\omega_{A}}}\right)^{\!2}\xi^{\parallel}(\mathbf{r}_{A},\omega)
\right].
\label{DOSparper}
\end{eqnarray}
Here, $\xi^{\perp(\parallel)}(\mathbf{r}_{A},\omega)$ is the
transverse (longitudinal) \emph{local} photonic DOS function
defined by~\cite{Bondarev04PRB}
\begin{equation}
\xi^{\perp(\parallel)}(\mathbf{r}_{A},\omega)=
{\Gamma^{\perp(\parallel)}(\mathbf{r}_{A},\omega)\over{\Gamma_{0}(\omega)}}\,,
\label{DOSdef}
\end{equation}
where
\begin{equation}
\Gamma^{\perp(\parallel)}(\mathbf{r}_{A},\omega)={8\pi\omega^{2}\over{\hbar
c^{2}}}\,d_{z}^{\,2}\,\mbox{Im}^{\perp(\parallel)}
G_{zz}^{\perp(\parallel)}(\mathbf{r}_{A},\mathbf{r}_{A},\omega)
\label{Gammaperpar}
\end{equation}
is the transverse (longitudinal) atomic spontaneous decay rate
near the CN with
\begin{eqnarray}
^{\perp(\parallel)}G_{zz}^{\perp(\parallel)}(\mathbf{r}_{A},\mathbf{r}_{A},\omega)=
\!\int\!d\mathbf{r}\,d\mathbf{r}^{\prime}\,\delta^{\perp(\parallel)}_{zz}
(\mathbf{r}_{A}-\mathbf{r})\nonumber\\
\times\;G_{zz}(\mathbf{r},\mathbf{r}^{\prime},\omega)\,
\delta^{\perp(\parallel)}_{zz}(\mathbf{r}^{\prime}-\mathbf{r}_{A})\mbox{\hskip1cm}
\label{Gzzperparperpar}
\end{eqnarray}
being the Green tensor $zz$-component transverse (longitudinal)
with respect to both variables, and $\Gamma_{0}(\omega)$ is the
same quantity in vacuum where~\cite{Abrikosov}
\begin{equation}
\mbox{Im}\,G^{0}_{zz}={\omega\over{6\pi c}}\,. \label{G0}
\end{equation}
Note that the transverse and longitudinal imaginary Green tensor
components in Eq.~(\ref{Gammaperpar}) can be written as
\begin{eqnarray}
\mbox{Im}^{\perp}G_{zz}^{\perp}&=&\mbox{Im}\,G^{0}_{zz}+\mbox{Im}^{\perp}
\overline{G}_{zz}^{\,\perp}\,,\label{ImGzzper}\\
\mbox{Im}^{\,\parallel}G_{zz}^{\parallel}&=&\mbox{Im}^{\,\parallel}
\overline{G}_{zz}^{\,\parallel} \label{ImGzzppar}
\end{eqnarray}
with $^{\perp(\parallel)}\overline{G}_{zz}^{\perp(\parallel)}$
representing the "pure" CN contribution to the total imaginary
Green tensor.~The longitudinal imaginary Green tensor in
Eq.~(\ref{ImGzzppar}) is totally contributed by a (longitudinal)
CN static polarization field and, therefore, does not contain the
vacuum term which is the transverse one by its definition as there
are no polarization Coulomb sources in vacuum.~Thus,
Eq.~(\ref{DOSdef}) may, in view of
Eqs.~(\ref{Gammaperpar})--(\ref{ImGzzppar}), be rewritten in the
form
\begin{eqnarray}
\xi^{\perp}(\mathbf{r}_{A},\omega)&=&
1+\overline{\xi}^{\perp}(\mathbf{r}_{A},\omega)\,,\label{xiper}\\
\xi^{\parallel}(\mathbf{r}_{A},\omega)&=&
\overline{\xi}^{\,\parallel}(\mathbf{r}_{A},\omega)\,,\label{xipar}
\end{eqnarray}
where the $\mathbf{r}_{A}$-dependent terms come totally from the
presence of the CN.

Eqs.~(\ref{Htwolev})-(\ref{gperppar}) represent the total secondly
quantized Hamiltonian of an atomic system interacting with the
electromagnetic field in the vicinity of the carbon nanotube.~In
deriving this Hamiltonian, there were only two standard
approximations done.~They are the electric dipole approximation
and the two-level approximation. The rotating wave approximation
commonly used was not applied, and the (diamagnetic)
$\hat{\mathbf{A}}^{\!2}$-term of the atom-field interaction
[Eq.~(\ref{Haf2})] was not neglected.~The latter two
approximations are justified in analyzing an atomic state
evolution due to \emph{real} dipole transitions to the nearest
states~\cite{Eberly}, described for two-level systems by the
$\mbox{g}^{(+)}$-terms of the Hamiltonian~(\ref{Hint}).~They are,
however, inappropriate in calculating the energy, where the
initial energy level is being shifted due to the perturbation
caused by \emph{virtual} dipole transitions via intermediate
atomic states, so that the system always remains in the initial
state and the terms neglected contribute to the level shift to the
same order of magnitude.~For two-level systems, this is described
by $\mbox{g}^{(-)}$-terms of the Hamiltonian~(\ref{Hint}) (usually
neglected within the rotating wave approximation) together with
the atomic transition frequency~(\ref{omegarenorm}) renormalized
due to the (usually discarded) $\hat{\mathbf{A}}^{\!2}$-term of
the atom-field interaction.

\begin{figure}[t]
\epsfxsize=8.65cm\centering{\epsfbox{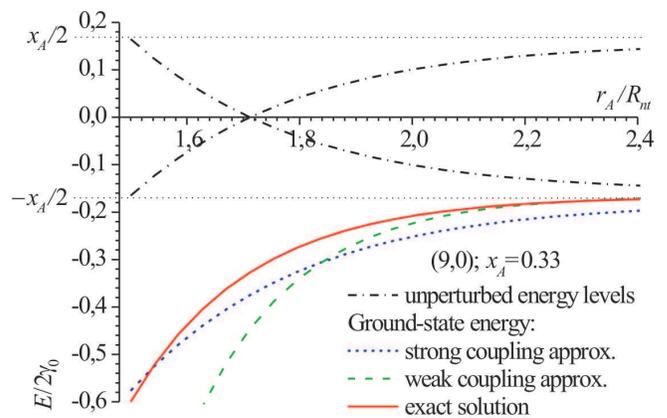}} \caption{(Color
online) Dimensionless unperturbed energy levels [given by the
Hamiltonian~(\ref{H0})] and total ground-state energy [given by
Eqs.~(\ref{Evdw})-(\ref{ksipar})] as functions of the atomic
position for the two-level atom outside the (9,0) CN.}
\label{fig2}
\end{figure}

\subsection{The van der Waals energy}\label{vdW}

The Hamiltonian~(\ref{Htwolev})-(\ref{Hint}) is a starting point
for the atom-nanotube vdW energy calculation.~Supposing the
two-level atom to be in the ground state, the vdW energy is
nothing but the $\mathbf{r}_{A}$-dependent contribution to the
ground-state eigenvalue of this Hamiltonian.~Important is that the
unperturbed atomic subsystem is now described by the Hamiltonian
(\ref{Harenorm}) with the renormalized transition frequency
(\ref{omegarenorm}) which is represented in terms of the local
photonic DOS (\ref{DOSdef}) as
\begin{equation}
\tilde{\omega}_{A}=\omega_{A}\!\left[1-{1\over{\pi}}\int_{0}^{\infty}\!\!\!d\omega
{\Gamma_{0}(\omega)\over{\omega^{2}}}\,\xi^{\perp}(\mathbf{r}_{A},\omega)\right].
\label{omegarenormxi}
\end{equation}
As is seen from Eq.~(\ref{omegarenormxi}), the transition
frequency $\tilde{\omega}_{A}$ decreases with increasing the
transverse local photonic DOS
$\xi^{\perp}(\mathbf{r}_{A},\omega)$, i.e.~when the atom
approaches the CN surface~\cite{Bondarev04PRB}, thereby bringing
the unperturbed atomic levels together, or even making them
degenerated if $\xi^{\perp}(\mathbf{r}_{A},\omega)$ is large
enough (see a typical example in Fig.~\ref{fig2}).~If one uses the
standard perturbation theory, this would yield the divergence of
high-order corrections to the vdW energy at small atom-CN-surface
distances because of the small energy denominators in the
corresponding power-series expansions.~To account for a possible
degeneracy of the unperturbed atomic levels in a correct way, one
has to use the perturbation theory for degenerated atomic levels
(see, e.g.,~\cite{Davydov}).~In so doing, one has to take the
upper state degeneracy into account of the whole system with
respect to $\mathbf{R}$, so that the ground-state wave function of
the whole system should be represented as a coherent mixture of
the lower and upper states of the form
\begin{equation}
|\psi\rangle=C_{l}\,|l\rangle|\{0\}\rangle+
\int_{0}^{\infty}\!\!\!\!\!d\omega\!\int\!d\mathbf{R}\,
C_{u}(\mathbf{R},\omega)\,|u\rangle|\{1(\mathbf{R},\omega)\}\rangle,
\label{wfunc}
\end{equation}
where $|\{0\}\rangle$ is the vacuum state of the field subsystem,
$|\{1(\mathbf{R},\omega)\}\rangle$ is its single-quantum excited
state, $C_{l}$ and $C_{u}(\mathbf{R},\omega)$ are unknown mixing
coefficients of the lower and upper states of the \emph{whole}
system.

The total energy $E$ of the ground state is now given by the
solution of a secular equation obtained by applying the
Hamiltonian~(\ref{Htwolev})-(\ref{Hint}) to the wave
function~(\ref{wfunc}).~This yields the integral equation
\begin{equation}
E=-{\hbar\tilde{\omega}_{A}\over{2}}-
\int_{0}^{\infty}\!\!\!\!\!d\omega\int\!d\mathbf{R}\,
{\displaystyle|\mbox{g}^{(-)}(\mathbf{r}_{A},\mathbf{R},\omega)|^{2}
\over{\displaystyle\hbar\omega+{\hbar\tilde{\omega}_{A}\over{2}}-E}}\,,
\label{E}
\end{equation}
which the ground-state-atom vdW energy $E_{vw}(\mathbf{r}_{A})$ is
determined from by means of an obvious relation
\begin{equation}
E=-{\hbar\omega_{A}\over{2}}+E_{vw}(\mathbf{r}_{A})\,,
\label{Evdw}
\end{equation}
with the first term representing the 'bare' (non-inte\-racting)
two-level atom in the ground-state.~Using dimensionless variables
\begin{equation}
\varepsilon_{vw}(\mathbf{r}_{A})={E_{vw}(\mathbf{r}_{A})\over{2\gamma_{0}}}\,,\hskip0.2cm
\tilde{\Gamma}_{0}(x)={\hbar\Gamma_{0}(x)\over{2\gamma_{0}}}\,,\hskip0.2cm
x={\hbar\omega\over{2\gamma_{0}}} \label{dimless}
\end{equation}
with $\gamma_{0}=2.7$~eV being the carbon nearest neighbor hopping
integral appearing in the CN surface axial conductivity
$\sigma_{zz}$ in Eq.~(\ref{currentCN}), and
Eqs.~(\ref{DOSparper}), (\ref{xiper})--(\ref{omegarenormxi}), one
can represent the dimensionless vdW energy
$\varepsilon_{vw}(\mathbf{r}_{A})$ in terms of the transverse and
longitudinal distance-dependent photonic DOS's in the vicinity of
the nanotube.~In so doing, one arrives at the integral equation of
the form
\begin{widetext}
\begin{equation}
\varepsilon_{vw}(\mathbf{r}_{A})={x_{A}\over{2\pi}}
\int_{0}^{\infty}\!\!\!dx{\tilde{\Gamma}_{0}(x)\over{x^{2}}}\,
\overline{\xi}^{\,\perp}(\mathbf{r}_{A},x)-
{x^{2}_{A}\over{2\pi}}\int_{0}^{\infty}\!\!\!dx{\tilde{\Gamma}_{0}(x)\over{x^{2}}}
{\displaystyle\overline{\xi}^{\,\perp}(\mathbf{r}_{A},x)+
{\left(x\over{x_{A}}\right)^{\!2}}\overline{\xi}^{\,\parallel}(\mathbf{r}_{A},x)
\over{\displaystyle
x+x_{A}\left[1-{1\over{2\pi}}\int_{0}^{\infty}\!\!\!dx
{\tilde{\Gamma}_{0}(x)\over{x^{2}}}\,\overline{\xi}^{\,\perp}(\mathbf{r}_{A},x)\right]
-\varepsilon_{vw}(\mathbf{r}_{A})}} \label{Evw}\vspace{0.75cm}
\end{equation}
with $\overline{\xi}^{\,\perp(\parallel)}(\mathbf{r}_{A},x)$ given
for $r_{A}\!>\!R_{cn}$ by (Appendix~\ref{appB})\vspace{0.5cm}
\begin{eqnarray}
\overline{\xi}^{\,\perp}(\mathbf{r}_{A},x)&=&{3\over{\pi}}\,\mbox{Im}\!\!\!
\sum_{p=-\infty}^{\infty}\int_{C}\!{dy\,s(R_{cn},x)v(y)^{4}
I_{p}^{2}[v(y)u(R_{cn})x]K_{p}^{2}[v(y)u(r_{A})x]\over{1+
s(R_{cn},x)v(y)^{2}I_{p}[v(y)u(R_{cn})x]K_{p}[v(y)u(R_{cn})x]}}\,,
\label{ksiper}\\[0.5cm]
\overline{\xi}^{\,\parallel}(\mathbf{r}_{A},x)&=&{3\over{\pi}}\,\mbox{Im}\!\!\!
\sum_{p=-\infty}^{\infty}\int_{C}\!{dy\,s(R_{cn},x)y^{2}v(y)^{2}
I_{p}^{2}[v(y)u(R_{cn})x]K_{p}^{2}[v(y)u(r_{A})x]\over{1+
s(R_{cn},x)v(y)^{2}I_{p}[v(y)u(R_{cn})x]K_{p}[v(y)u(R_{cn})x]}}\,,
\label{ksipar}\vspace{0.75cm}
\end{eqnarray}
\end{widetext}
where $I_{p}$ and $K_{p}$ are the modified cylindric Bessel
functions, $v(y)\!=\!\sqrt{y^{2}-1}\,$,
$u(r)\!=\!2\gamma_{0}r/\hbar c$, and $s(R_{cn},x)\!=\!2i\alpha
u(R_{cn})x\overline{\sigma}_{zz}(R_{cn},x)$ with
$\overline{\sigma}_{zz}\!=\!2\pi\hbar\sigma_{zz}/e^{2}$ being the
dimensionless CN surface conductivity per unit length and
$\alpha\!=\!e^{2}/\hbar c\!=\!1/137$ representing the
fine-structure constant.~The integration contour $C$ goes along
the real axis of the complex plane and envelopes the branch points
$y\!=\!\pm 1$ of the function $v(y)$ in the integrands from below
and from above, respectively.~For $r_{A}\!<\!R_{cn}$,
Eqs.~(\ref{ksiper}) and (\ref{ksipar}) are modified by a simple
replacement $r_{A}\!\leftrightarrow\!R_{cn}$ in the Bessel
function arguments in the numerators of the
integrands.

\section{Qualitative analysis}\label{qualitatively}

Eqs.~(\ref{Evw})-(\ref{ksipar}) describe the ground-state-atom vdW
energy near an infinitely long single-wall carbon nanotube in
terms of the local (distance-dependent) photonic
DOS.~Eq.~(\ref{Evw}) is universal in the sense that it covers both
strong and weak atom-field coupling regimes defined as those
violating and non-violating, respectively, the applicability
domain of the conventional stationary perturbation
theory~\footnote{One should make a clear difference between this
definition and that where the strong (weak) atom-field coupling
regime is defined as that violating (non-violating) the
applicability domain of the Markovian
approximation~\cite{John}.~Our definition here is based on the
Born approximation associated with the strength of the coupling
between the atom and the photonic reservoir, whereas the Markovian
approximation is related to memory effects of the photonic
reservoir.}.

\subsubsection{Weak coupling regime}

If, when the atom is faraway from the CN, the local photonic DOS
$\overline{\xi}^{\,\perp}(\mathbf{r}_{A},x)$ is such small that
\begin{equation}
{1\over{2\pi}}\int_{0}^{\infty}\!\!\!dx{\tilde{\Gamma}_{0}(x)\over{x^{2}}}\,
\overline{\xi}^{\,\perp}(\mathbf{r}_{A},x)\ll1 \label{DOSweakcond}
\end{equation}
and
\begin{equation}
\varepsilon_{vw}(\mathbf{r}_{A})\sim0\,,
\label{vdwweakcond}
\end{equation}
then Eq.~(\ref{Evw}) yields a well-known perturbation theory
result (see, e.g.,~\cite{Buhmann04JOB}) of the form
\begin{eqnarray}
\varepsilon_{vw}(\mathbf{r}_{A})\approx{x_{A}\over{2\pi}}
\int_{0}^{\infty}\!\!\!dx{\tilde{\Gamma}_{0}(x)\over{x^{2}}}\,
\overline{\xi}^{\,\perp}(\mathbf{r}_{A},x)\nonumber\\
-{x^{2}_{A}\over{2\pi}}\int_{0}^{\infty}\!\!\!dx{\tilde{\Gamma}_{0}(x)\over{x^{2}}}
{\displaystyle\overline{\xi}^{\,\perp}(\mathbf{r}_{A},x)+
{\left(x\over{x_{A}}\right)^{\!2}}\overline{\xi}^{\,\parallel}(\mathbf{r}_{A},x)
\over{\displaystyle x+x_{A}}}\,, \label{Evwweak}
\end{eqnarray}
where the first term comes from the unperturbed Hamiltonian
(\ref{H0}),~(\ref{Harenorm}) and the second one is the second
order correction due to the perturbation (\ref{Hint}).~This can be
equivalently rewritten in the form
\begin{eqnarray}
\varepsilon_{vw}(\mathbf{r}_{A})&\approx&{x_{A}\over{2\pi}}
\int_{0}^{\infty}\!\!\!dx{\tilde{\Gamma}_{0}(x)\over{x(x+x_{A})}}\nonumber\\
&\times&\left[\overline{\xi}^{\,\perp}(\mathbf{r}_{A},x)-
{x\over{x_{A}}}\,\overline{\xi}^{\,\parallel}(\mathbf{r}_{A},x)\right],
\label{Evwweak1}
\end{eqnarray}
from which, in view of Eqs.~(\ref{ksiper}),~(\ref{ksipar}) and
basic properties of the modified cylindric Bessel
functions~\cite{Abramovitz}, one can immediately come to an
interesting conclusion.~Namely, if the atom is fixed outside
(inside) the CN in a way that the weak atom-field coupling regime
is realized, i.e.~far enough from the CN surface, then the modulus
of the atom-nanotube vdW energy increases (decreases) with the CN
radius.~The conclusion is physically clear as the effective
atom-nanotube interaction area is larger (smaller) for
larger-radius nanotubes when the atom is outside (inside) the
CN.~Important, however, is that this obvious conclusion is,
strictly speaking, only valid in the weak atom-field coupling
regime for the outside atomic position, while for the inside
atomic position it represents a general effect of the effective
interaction area reduction with lowering the CN surface curvature.

In the large CN radius limit, Eq.~(\ref{Evwweak1}) along with
Eqs.~(\ref{ksiper}) and (\ref{ksipar}) can be shown to reproduce a
well-known Casimir-Polder result~\cite{Casimir} for an atom near
an infinitely conducting plane (see Appendix~\ref{appC}).

\subsubsection{Strong coupling regime}

When the atom is close enough to the CN surface, the local
photonic DOS $\overline{\xi}^{\,\perp}(\mathbf{r}_{A},x)$ is
large, so that one might expect the condition
\begin{equation}
{1\over{2\pi}}\int_{0}^{\infty}\!\!\!dx{\tilde{\Gamma}_{0}(x)\over{x^{2}}}\,
\overline{\xi}^{\,\perp}(\mathbf{r}_{A},x)\sim1
\label{DOSstrongcond}
\end{equation}
to hold true.~Under this condition, Eq.~(\ref{Evw}) reduces to an
integral equation of the form
\begin{eqnarray}
\varepsilon_{vw}(\mathbf{r}_{A})\approx{x_{A}\over{2\pi}}
\int_{0}^{\infty}\!\!\!dx{\tilde{\Gamma}_{0}(x)\over{x^{2}}}\,
\overline{\xi}^{\,\perp}(\mathbf{r}_{A},x)\nonumber\\
-{x^{2}_{A}\over{2\pi}}\int_{0}^{\infty}\!\!\!dx{\tilde{\Gamma}_{0}(x)\over{x^{2}}}
{\displaystyle\overline{\xi}^{\,\perp}(\mathbf{r}_{A},x)+
{\left(x\over{x_{A}}\right)^{\!2}}\overline{\xi}^{\,\parallel}(\mathbf{r}_{A},x)
\over{\displaystyle x-\varepsilon_{vw}(\mathbf{r}_{A})}}
\label{Evwstrong}
\end{eqnarray}
valid in the strong atom-field coupling regime.

Eq.~(\ref{DOSstrongcond}) corresponds to the \emph{inverted}
unperturbed atomic levels described by the Hamiltonian
(\ref{Harenorm}) with $\tilde{\omega}_{A}$ given by
Eq.~(\ref{omegarenormxi}) (see Fig.~\ref{fig2} as an example). The
levels are degenerated when the left-hand side of
Eq.~(\ref{DOSstrongcond}) is exactly $0.5$.~A noteworthy point is
that if one used the weak coupling approximation of
Eq.~(\ref{Evw}) for the atom close to the CN surface where
Eq.~(\ref{DOSweakcond}) is not satisfied and
Eq.~(\ref{DOSstrongcond}) holds true instead, then the result
would be divergent at the lower limit of integration.~To avoid
this fact, one has to use either strong coupling approximation
(\ref{Evwstrong}) or the exact equation~(\ref{Evw}) to calculate
the vdW energy of the atom close to the nanotube surface.

\section{Numerical results and discussion}\label{numerically}

Using Eqs.~(\ref{Evdw})-(\ref{ksipar}), we have simulated the
total ground state energy of the whole system "two-level
atom~+~CN-modified vacuum electromagnetic field" and the vdW
energy of the two-level atom nearby metallic and semiconducting
carbon nanotubes of different radii.~The dimensionless free-space
spontaneous decay rate in Eq.~(\ref{Evw}) was approximated by the
expression
$\tilde{\Gamma}_{0}(x)\approx4\alpha^{3}x^{3}/3x^{2}_{A}$
($\alpha=1/137$ is the fine-structure constant) valid for atomic
systems with Coulomb interaction~\cite{Davydov}.

Figure~\ref{fig2} shows the dimensionless ground state energy
level of the whole system [given by the corresponding eigenvalue
of the the total Hamiltonian~(\ref{Htot}) and represented by
Eqs.~(\ref{Evdw})-(\ref{ksipar})] and two dimensionless energy
levels of the unperturbed Hamiltonian~(\ref{H0}),~(\ref{Harenorm})
as functions of the atomic position outside the (9,0) CN.~As the
atom approaches the CN surface, its unperturbed levels come
together, then get degenerated and even inverted at a very small
atom-surface distance.~In so doing, the weak coupling
approximation for the ground state energy [given by
Eqs.~(\ref{Evdw}),~(\ref{dimless}),~(\ref{Evwweak1}) and
(\ref{ksiper}),~(\ref{ksipar})] diverges near the surface, whereas
the strong coupling approximation
[Eqs.~(\ref{Evdw}),~(\ref{dimless}),~(\ref{Evwstrong}) and
(\ref{ksiper}),~(\ref{ksipar})] yields a finite result.~The exact
solution reproduces the weak-coupling approximation at large and
the strong-coupling approximation at short atom-surface distances,
respectively.

\begin{figure}[t]
\epsfxsize=8.65cm\centering{\epsfbox{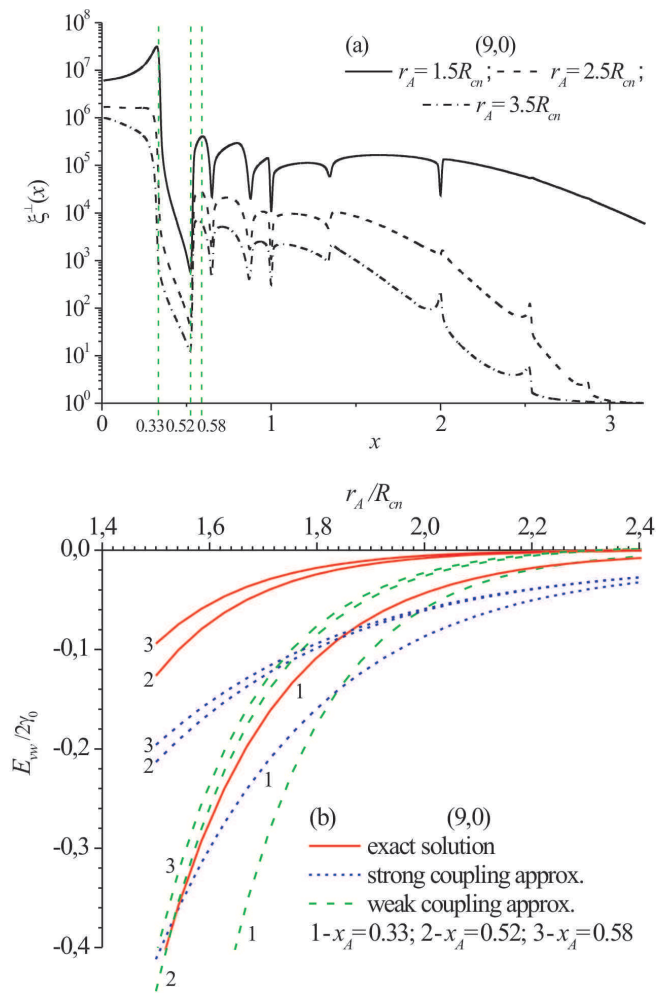}} \caption{(Color
online) Transverse local photonic DOS's (a) and ground-state vdW
energies (b) as functions of the atomic position for the two-level
atom outside the (9,0) CN. The 'bare' atomic transition
frequencies are indicated by dashed lines in Fig.~\ref{fig3}(a).}
\label{fig3}
\end{figure}

\begin{figure}[b]
\epsfxsize=8.65cm\centering{\epsfbox{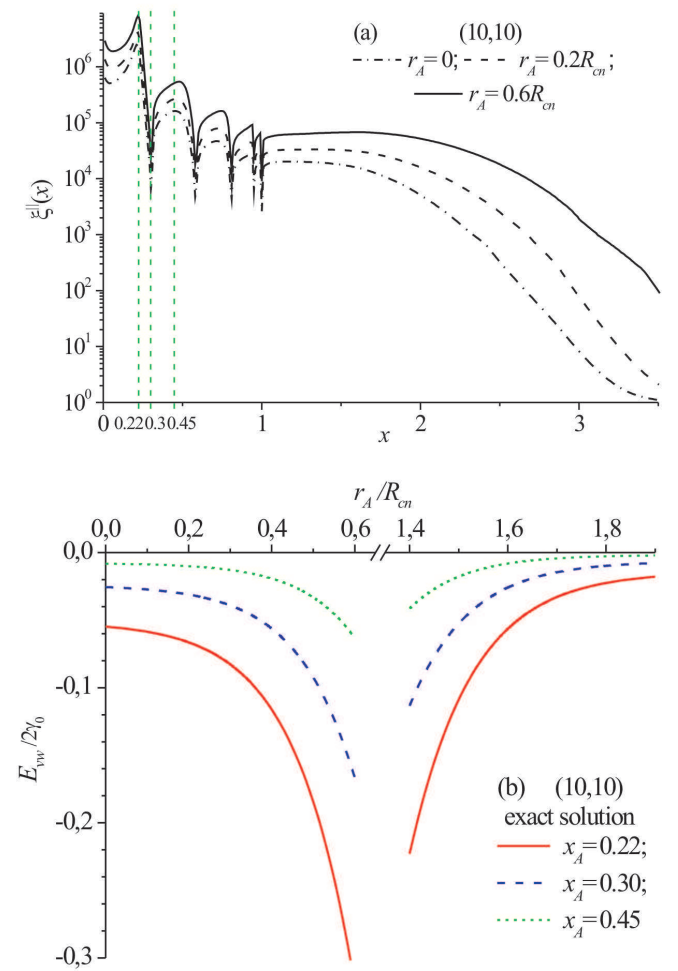}} \caption{(Color
online) Longitudinal local photonic DOS's (a) and ground-state vdW
energies (b) as functions of the atomic position for the two-level
atom nearby the (10,10) CN. The 'bare' atomic transition
frequencies are indicated by dashed lines in Fig.~\ref{fig4}(a).}
\label{fig4}
\end{figure}

The degeneracy of the unperturbed atomic levels is the consequence
of the transverse local photonic DOS increase [see
Eq.~(\ref{omegarenormxi})] as the atom approaches the CN
surface.~A typical example is shown in Fig.~\ref{fig3}~(a) for the
atom outside the (9,0) CN.~The transverse DOS function is seen to
increase with decreasing the atom-CN-surface distance,
representing the atom-field coupling strength enhancement with
surface photonic modes of the nanotube as the atom approaches the
nanotube surface~\cite{Bondarev02,Bondarev04PRB}.~The vertical
dashed lines indicate the 'bare' atomic transition frequencies
$x_{A}$ for which the vdW energies shown in Fig.~\ref{fig3}~(b)
were calculated.~Some of them are the positions of peaks, the
others are the positions of dips of the local photonic DOS.~They
are typical for some rear-earth ions such as Europium, or for
heavy hydrogen-like atoms such as Caesium (supposed to be
non-ionized near the CN). For example~\cite{Schniepp}, for
a~well-known two-level dipole transition
$^{7}F_{2}\leftrightarrow\,\!^{5}D_{0}$ of Europium at
$\lambda=615$~nm one has $x_{A}\approx0.37$, whereas for Caesium
one obtains from its first ionization potential~\cite{Lide} the
estimate
$x_{A}\approx3.89\mbox{\,eV}\times3/4\times(2\gamma_{0})^{-1}\approx0.5$
(the factor $3/4$ comes from the Lyman series of~Hydrogen), or
less for highly excited Rydberg states.

Figure~\ref{fig3}~(b) shows the vdW energies of the atom outside
the (9,0) CN for three different 'bare' atomic transition
frequencies [indicated by vertical dashed lines in
Fig.~\ref{fig3}~(a)].~Here, exact numerical solutions of
Eqs.~(\ref{Evw})-(\ref{ksipar}) are compared with those obtained
within the weak coupling approximation
(\ref{Evwweak1}),~(\ref{ksiper}),~(\ref{ksipar}) and within the
strong coupling approximation (\ref{Evwstrong}),
(\ref{ksiper}),~(\ref{ksipar}).~At small atom-CN-surface
distances, the exact solutions are seen to be fairly well
reproduced by those obtained in the strong coupling approximation,
clearly indicating the strong atom-field coupling regime in a
close vicinity of the nanotube surface.~The deviation from the
strong coupling approximation increases with transition
frequency~$x_{A}$, that is easily explicable since the degeneracy
condition (\ref{DOSstrongcond}) of the unperturbed atomic levels
is more difficult to reach for larger inter-level separations.~As
the atom moves away from the CN surface, the exact solutions
deviate from the strong coupling solutions and approach those
given by the weak coupling approximation, indicating the reduction
of the atom-field coupling strength with raising the atom-surface
distance.~The weak-coupling solutions are seen to be divergent
close to the nanotube surface as it should be because of the
degeneracy of the unperturbed atomic levels in this region.

Figure~\ref{fig4}~(a) shows the longitudinal local photonic DOS's
for the atom inside the (10,10) CN at three representative
distances from the nanotube wall.~As the atom approaches the wall,
the longitudinal DOS function increases in a similar way as the
transverse DOS function does in Fig.~\ref{fig3}~(a).~The vertical
dashed lines indicate the 'bare' atomic transition frequencies for
which the atomic vdW energies in Fig.~\ref{fig4}~(b) were
calculated.~Figure~\ref{fig4}~(b) represents the exact solutions
of Eq.~(\ref{Evw}) for atom inside and outside the (10,10)
CN.~When the atom is inside, the vdW energies are seen to be in
general lower than those for the atom outside the CN.~This may be
attributed to the fact that, due to the nanotube curvature, the
effective interaction area between the atom and the CN surface is
larger when the atom is inside rather than when it is outside the
CN.~This, in turn, indicates that encapsulation of doped atoms
into the nanotube is energetically more favorable than their
outside adsorption by the nanotube surface --- the effect observed
experimentally in Ref.~\cite{Jeong}.

Comparing Fig.~\ref{fig3}~(b) with Fig.~\ref{fig4}~(b) for the
outside atomic position, one can see that, if the atom-CN-surface
distance is so large that the weak-coupling approximation is good,
then for the same distance and approximately the same atomic
transition frequency $x_{A}\approx0.3$ the vdW energy near the
(10,10) nanotube ($R_{cn}=6.78$~\AA) is lower than that near the
(9,0) nanotube ($R_{cn}=3.52$~\AA).~This is in agreement with the
qualitative predictions formulated above in analyzing the weak
coupling regime.~In the present case, the effective
atom-CN-surface interaction area is larger for the atom outside
the larger-radius (10,10) nanotube, thereby explaining the
effect.~Obviously, one should expect an opposite effect for the
inside atomic position.~Experimentally, the heat of absorbtion of
Krypton, to take an example, has been shown to be larger for
graphite (0.17~eV/atom) than for multiwall nanotubes
(0.12~eV/atom)~\cite{Masenelli}, in agreement with our theoretical
predictions.

\begin{figure}[t]
\epsfxsize=8.65cm\centering{\epsfbox{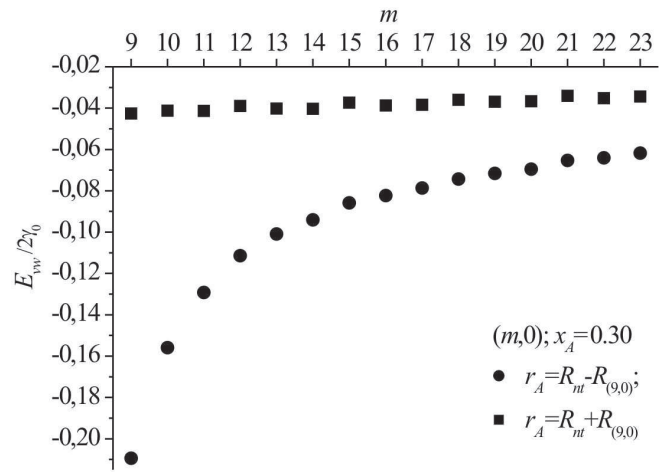}}
\caption{Ground-state van der Waals energy of the two-level atom
positioned at a fixed atom-surface distance [equal to the radius
of the (9,0) CN] inside and outside "zigzag" $(m,0)$ CNs of
different radii, as a function of the nanotube index $m$.}
\label{fig5}
\end{figure}

A similar tendency is demonstrated in Fig~\ref{fig5}.~Here, the
vdW energies calculated from Eqs.~(\ref{Evw})-(\ref{ksipar}) are
represented for the atom positioned at a fixed distance from the
nanotube wall [chosen to be equal to the radius of the (9,0) CN]
inside and outside the "zigzag" $(m,0)$ CNs of increasing radius
$R_{cn}=m\,(a\sqrt{3}/2\pi)$ ($a\!=\!1.42$~\AA~is the CN
interatomic distance).~When the atom is inside, the energy
absolute value goes down with $R_{cn}$, representing a general
tendency of the effective interaction area reduction with lowering
the CN surface curvature.~When the atom is outside, the effect
would be the opposite one if the atom were weakly coupled to the
field, thus indicating that the atom-field coupling regime is not
actually weak for the distance chosen.~In general, the inside and
outside vdW energies approach each other with $R_{cn}$ as it
should be since the two atomic positions become equivalent in the
plane limit where $R_{cn}\!\rightarrow\!\infty$.~Interesting is
also the fact that the vdW energies of the atom outside the
metallic nanotubes ($m$ is divisible by 3) are on average a little
bit smaller than those for the atom outside semiconducting
ones.~This is certainly the property related to the difference in
the conductivities of metallic and semiconducting "zigzag"
nanotubes.

\section{Conclusions}\label{conclusion}

Recently, a theory of atomic spontaneous decay near CNs was
developed~\cite{Bondarev02,Bondarev04PLA,Bondarev04PRB}.~The
theory has brought out fascinating peculiarities of the
vacuum-field interactions in atomically doped CNs.~In
particular~\cite{Bondarev04PLA,Bondarev04PRB}, if the atom is
close enough to the CN surface and the atomic transition frequency
is in the vicinity of a resonance of the photonic DOS, the atomic
spontaneous decay dynamics exhibits vacuum-field Rabi
oscillations.~This is a principal signature of strong coupling of
the excited atomic state to vacuum surface photonic modes in the
CN.~Such a strongly coupled atomic state is nothing but a
'quasi-1D cavity polariton' similar to quasi-0D excitonic
(electronic) polaritons in quantum dot microcavities widely
discussed in the literature~\cite{Weisbuch,Gerard,Dini}.~The study
of such phenomena was started awhile ago in atomic
physics~\cite{Haroche1} and still attracts a lot of interest
especially in connection with research on quantum
computation~\cite{Haroche2}.

Obviously, the stability of quasi-1D atomic polaritons in CNs is
mainly determined by the atom-nanotube van der Waals
interaction.~This interaction, however, originates from
\emph{strong} atom-field coupling and, therefore, cannot be
correctly described in terms of vacuum-QED-based (weak-coupling)
vdW interaction models as well as in terms of those based upon the
linear response theory.~In the present paper, we have developed a
simple quantum mechanical approach to the ground-state vdW energy
calculation of a (two-level) atomic system near a CN.~The approach
is based upon the perturbation theory for degenerated atomic
levels, thus accounting for both weak and strong atom-field
coupling and thereby covering the vdW interactions of the
ground-state quasi-1D atomic polaritons in CNs.~Within the
framework of this approach, the vdW energy is described by the
integral equation represented in terms of the local photonic
DOS.~By solving this equation numerically, we have demonstrated
the inapplicability of weak-coupling-based vdW interaction models
in a close vicinity of the nanotube surface where the local
photonic DOS effectively increases, giving rise to an atom-field
coupling enhancement followed by the degeneracy of the unperturbed
atomic levels due to the diamagnetic interaction term.

The fundamental conclusion above is supplemented by those
important for various applications of atomically doped CNs in
modern nanotechnology.~In particular, we have studied CN surface
curvature effects on the atom-nanotube vdW interactions and have
shown that an inside encapsulation of doped atoms into the
nanotube is energetically more favorable than their outside
adsorption by the nanotube surface, in agreement with the
experimental observations reported in Ref.~\cite{Jeong}.~Moreover,
if the atom is fixed outside the CN in a way that the weak
atom-field coupling regime is realized, i.e.~far enough from the
CN surface, then the modulus of the atom-nanotube vdW energy
increases with the CN radius because the effective atom-nanotube
interaction area is larger for larger-radius nanotubes in this
case.~For inside atomic position, the modulus of the atom-nanotube
vdW energy decreases with the CN radius, representing a general
effect of the effective interaction area reduction with lowering
the CN surface curvature.

A further intriguing extension of the present work could be the
vdW interactions of excited atomic states where, even in the weak
atom-field-coupling regime and in the simplest case of an atom
near a planar semi-infinite medium, very interesting peculiarities
(e.g., oscillatory behavior) were recently shown to
exist~\cite{Buhmann04pre}.

\acknowledgments

Discussions with Prof. D.-G.Welsch and Dr. S.Y.Buh\-mann are
gratefully acknowledged.~The work was done in part within the
framework of the Belgian PAI-P5/01 project during the stay of one
of the authors (I.B.) in the University of Namur,
Belgium.~I.B.~thanks the Belgian OSTC.

\appendix

\section{The two-level approximation}\label{appA}

In this Appendix, we rewrite the
Hamiltonian~(\ref{Htot})-(\ref{Haf2}) in terms of a two-level
atomic model~\cite{Eberly}.~Within the framework of this model,
the spectrum of the atomic Hamiltonian~(\ref{Ha}) is approximated
by the two eigenstates, upper $|u\rangle$ and lower $|l\rangle$,
with the energies $\hbar\omega_{u}$ and $\hbar\omega_{l}$,
respectively.~One has then
\begin{equation}
\hat{H}_{A}\approx\hbar\omega_{u}|u\rangle\langle u|+
\hbar\omega_{l}|l\rangle\langle l| \label{Haapp1}
\end{equation}
with a completeness relation
\begin{equation}
|u\rangle\langle u|+|l\rangle\langle l|=\hat{I}.
\label{completeness}
\end{equation}
Subtracting a constant term
$(\hbar/2)(\omega_{u}+\omega_{l})\hat{I}$ from the
Hamiltonian~(\ref{Haapp1}) and thereby placing the energy zero in
the middle between the two energy levels, one arrives at the
'bare' two-level atomic Hamiltonian of the form
\begin{equation}
\hat{H}_{A}={\hbar\omega_{A}\over{2}}\,\hat{\sigma}_{z}
\label{Hbare}
\end{equation}
with $\hat{\sigma}_{z}=|u\rangle\langle u|-|l\rangle\langle l|$
and $\omega_{A}=\omega_{u}-\omega_{l}$ being the 'bare' atomic
transition frequency.

In terms of such a~two-level scheme, the atomic dipole moment
operator $\hat{\mathbf{d}}$ has the matrix elements $\langle
u|\hat{\mathbf{d}}|l\rangle=\langle
l|\hat{\mathbf{d}}|u\rangle=\mathbf{d}$ and $\langle
u|\hat{\mathbf{d}}|u\rangle=\langle
l|\hat{\mathbf{d}}|l\rangle=0$.~Keeping this and the completeness
relation~(\ref{completeness}) in mind, the interaction
Hamiltonians~(\ref{Haf1}) and (\ref{Haf2}) can be rewritten as
follows.

For the interaction~(\ref{Haf1}), using Eqs.~(\ref{vecpot}) and
(\ref{scalpot}) and a~well-known quantum mechanical operator
equality
\begin{equation}
(\hat{\mathbf{p}}_{i})_{\alpha}=m_{i}{d\over{dt}}(\hat{\mathbf{r}}_{i})_{\alpha}
={m_{i}\over{i\hbar}}\,[(\hat{\mathbf{r}}_{i})_{\alpha},\hat{H}_{A}]\,,\\[0.25cm]
\label{p}
\end{equation}
where the Latin and Greek indexes enumerate particles in the atom
and vector components, respectively, and $\hat{H}_{A}$ is given by
Eq.~(\ref{Hbare}), one has
\begin{eqnarray}
\hat{H}^{(1)}_{AF}&=&(\hat{\sigma}-\hat{\sigma}^{\dag})\,\mathbf{d}\cdot\!\left[
\int_{0}^{\infty}\!\!\!\!\!d\omega{\omega_{A}\over{\omega}}\,
\underline{\hat{\mathbf{E}}}^{\perp}(\mathbf{r}_{A},\omega)-\mbox{h.c.}\right]\nonumber\\
&-&(\hat{\sigma}+\hat{\sigma}^{\dag})\,\mathbf{d}\cdot\!\left[
\int_{0}^{\infty}\!\!\!\!\!d\omega\,
\underline{\hat{\mathbf{E}}}^{\parallel}(\mathbf{r}_{A},\omega)+\mbox{h.c.}\right],
\label{Haf1app}
\end{eqnarray}
with $\hat{\sigma}=|l\rangle\langle u|$ and
$\hat{\sigma}^{\dag}=|u\rangle\langle l|$.~Using further
Eqs.~(\ref{elperpar}), (\ref{Erw}) and (\ref{currentCN}) and
assuming the longitudinal (along the CN axis) atomic dipole
orientation due to the dominant axial polarizability of the
nanotube~\cite{Tasaki,Li,Jorio,Marinop}, one arrives at the
secondly quantized interaction Hamiltonian~(\ref{Hint}) with the
interaction matrix elements (\ref{gpm}) and (\ref{gperppar}).

For the interaction~(\ref{Haf2}), one may proceed as
follows
\begin{eqnarray}
\hat{H}^{(2)}_{AF}=\sum_{i,j}\sum_{\alpha,\beta}{q_{i}q_{j}\over{2m_{i}c^{2}}}
\,\delta_{ij}\delta_{\alpha\beta}
\hat{A}_{\alpha}(\mathbf{r}_{A})\hat{A}_{\beta}(\mathbf{r}_{A})\nonumber\\
={i\over{2\hbar c^{2}}}\sum_{i,\alpha,\beta}{q_{i}\over{m_{i}}}
[(\hat{\mathbf{p}}_{i})_{\alpha},\hat{d}_{\beta}]
\hat{A}_{\alpha}(\mathbf{r}_{A})\hat{A}_{\beta}(\mathbf{r}_{A})\,,
\label{Haf2app1}
\end{eqnarray}
where the product of the two Kronecker-symbols in the first line
is represented in terms of the 'coordinate-momentum' commutator as
$$\delta_{ij}\delta_{\alpha\beta}={i\over{\hbar}}\,
[(\hat{\mathbf{p}}_{i})_{\alpha},(\hat{\mathbf{r}}_{j})_{\beta}].$$
Using Eq.~(\ref{p}) for the momentum operator components and the
completeness relation~(\ref{completeness}), one arrives at
\begin{eqnarray}
\hat{H}^{(2)}_{AF}=-{\omega_{A}\over{\hbar c^{2}}}\,
\hat{\sigma}_{z}\!\left[\mathbf{d}\cdot
\hat{\mathbf{A}}(\mathbf{r}_{A})\right]^{2}
\label{Haf2app2}
\end{eqnarray}
which, upon substituting Eq.~(\ref{vecpot}) for the vector
potential operator, becomes
\begin{widetext}
\[
-{\hat{\sigma}_{z}\over{\hbar\omega_{A}}}\sum_{\alpha,\beta}d_{\alpha}d_{\beta}
\left\{\int_{0}^{\infty}\!\!\!\!\!d\omega{\omega_{A}\over{i\omega}}\!
\int_{0}^{\infty}\!\!\!\!\!d\omega^{\prime}{\omega_{A}\over{i\omega^{\prime}}}\,
\underline{\hat{E}\!}_{\,\alpha}^{\perp}(\mathbf{r}_{A},\omega)\,
\underline{\hat{E}\!}_{\,\beta}^{\perp}(\mathbf{r}_{A},\omega^{\prime})
-\!\int_{0}^{\infty}\!\!\!\!\!d\omega{\omega_{A}\over{i\omega}}\!
\int_{0}^{\infty}\!\!\!\!\!d\omega^{\prime}{\omega_{A}\over{i\omega^{\prime}}}\,
\underline{\hat{E}\!}_{\,\alpha}^{\perp}(\mathbf{r}_{A},\omega)\!
\left[\underline{\hat{E}\!}_{\,\beta}^{\perp}(\mathbf{r}_{A},\omega^{\prime})\right]^{\dag}
\!+\mbox{h.c.}\right\}.
\]
\end{widetext}
The four items here can, in view of Eqs.~(\ref{elperpar}),
(\ref{Erw}) and (\ref{currentCN}), be classified as follows.~The
first item and its hermitian conjugate describe the processes with
simultaneous annihilation and creation of \emph{two} photons
(two-photon transitions).~They may be safely neglected if the
field intensity is low enough that, we believe, is the case for
the vacuum electromagnetic field we deal with in considering the
vdW interactions.~Furthermore, keeping them in the Hamiltonian
$\hat{H}^{(2)}_{AF}$ is meaningless as the Hamiltonian
$\hat{H}^{(1)}_{AF}$ is nothing but a dipole approximation which,
by its definition, neglects two-photon transitions.~The second
item and its hermitian conjugate are combined together by using
bosonic commutation relations (\ref{commut}) to give in terms of
the bosonic field operators and the transverse dipole interaction
matrix element~(\ref{gperppar}) the expression
\begin{widetext}
\[
-{\hat{\sigma}_{z}\over{\hbar\omega_{A}}}\int_{0}^{\infty}\!\!\!\!\!d\omega\!
\int\!d\mathbf{R}\,|\mbox{g}^{\perp}(\mathbf{r}_{A},\mathbf{R},\omega)|^{2}
-{2\hat{\sigma}_{z}\over{\hbar\omega_{A}}}\int_{0}^{\infty}\!\!\!\!\!d\omega\,
d\omega^{\prime}\!\int\!d\mathbf{R}\,d\mathbf{R}^{\prime}
\left[\mbox{g}^{\perp}(\mathbf{r}_{A},\mathbf{R},\omega)\right]^{\ast}\!
\mbox{g}^{\perp}(\mathbf{r}_{A},\mathbf{R}^{\prime},\omega^{\prime})
f^{\dag}(\mathbf{R},\omega)f(\mathbf{R}^{\prime},\omega^{\prime}),
\]
\end{widetext}
where the second term is nothing but a two-photon correction to
the dipole interaction~(\ref{Hint}), that can be easily seen by
noting that
$\hat{\sigma}_{z}=\hat{\sigma}^{\dag}\hat{\sigma}-\hat{\sigma}\hat{\sigma}^{\dag}$.~This
correction must be neglected for the reasons just discussed, so
that finally one arrives at
\begin{eqnarray}
\hat{H}^{(2)}_{AF}\approx-{\hat{\sigma}_{z}\over{\hbar\omega_{A}}}
\int_{0}^{\infty}\!\!\!\!\!d\omega\!\int\!d\mathbf{R}\,
|\mbox{g}^{\perp}(\mathbf{r}_{A},\mathbf{R},\omega)|^{2}.
\label{Haf2app3}
\end{eqnarray}
Eq.~(\ref{Haf2app3}) can now be combined with the 'bare' atomic
Hamiltonian~(\ref{Hbare}) to yield the 'effective' unperturbed
atomic Hamiltonian~(\ref{Harenorm}) with the renormalized atomic
transition frequency given by Eq.~(\ref{omegarenorm}).

Thus, we have proved that the
Hamiltonian~(\ref{Htot})-(\ref{Haf2}), being approximated in terms
of the two-level atomic model, is represented by the
Hamiltonian~(\ref{Htwolev})-(\ref{Hint}).

\section{The transverse and longitudinal local photonic DOS}\label{appB}

We start with Eq.~(\ref{GreenequCN}) for the Green tensor of the
electromagnetic subsystem.~This equation is a direct consequence
of the equation
\begin{equation}
\sum_{\alpha=r,\varphi,z}\!\!\!
\left(\bm{\nabla}\!\times\bm{\nabla}\!\times-\,k^{2}\right)_{\!z\alpha}
\underline{\hat{E}\!}_{\,\alpha}(\mathbf{r},\omega)=
i{4\pi\over{c}}\,k\,\underline{\hat{I}\!}_{\,z}(\mathbf{r},\omega)
\label{EIapp}
\end{equation}
obtained by substituting the magnetic field operator from
Eq.~(\ref{MaxwelE}) into Eq.~(\ref{MaxwelH}).~On the other hand,
under the condition of the Coulomb gauge
(see,~e.g.,~\cite{Jackson}), one has
\begin{equation}
\underline{\hat{\mathbf{E}}}(\mathbf{r},\omega)=
\underline{\hat{\mathbf{E}}}^{\perp}(\mathbf{r},\omega)+
\underline{\hat{\mathbf{E}}}^{\parallel}(\mathbf{r},\omega)
\label{EAPhiapp}
\end{equation}
with [compare with Eqs.~(\ref{vecpot}) and (\ref{scalpot})]
\begin{equation}
\underline{\hat{\mathbf{E}}}^{\perp}(\mathbf{r},\omega)=
ik\underline{\hat{\mathbf{A}}}(\mathbf{r},\omega)\,,
\label{Eperpapp}
\end{equation}
where
\begin{equation}
\bm{\nabla}\cdot\underline{\hat{\mathbf{A}}}(\mathbf{r},\omega)=0\,,
\label{Cgaugepapp}
\end{equation}
and
\begin{equation}
\underline{\hat{\mathbf{E}}}^{\parallel}(\mathbf{r},\omega)=
-\bm{\nabla}\varphi(\mathbf{r},\omega)\,.
\label{Eparapp}
\end{equation}
From Eqs.~(\ref{EIapp})--(\ref{Eparapp}), in view of
Eq.~(\ref{elperpar}) and (\ref{Erw}), one obtains the following
equation for the Green tensor components
\begin{eqnarray}
\sum_{\alpha=r,\varphi,z}\!\!\!
\left(\bm{\nabla}\!\times\bm{\nabla}\!\times-\,k^{2}\right)_{\!z\alpha}&&\mbox{\hskip-0.8cm}
\left[\;^{\!\!\perp}G_{\alpha
z}(\mathbf{r},\mathbf{R},\omega)+\,^{\!\!\parallel}G_{\alpha
z}(\mathbf{r},\mathbf{R},\omega)\right]\nonumber\\
&=&\delta(\mathbf{r}-\mathbf{R})
\label{Greenperparapp}
\end{eqnarray}
with additional constraints
\begin{equation}
\sum_{\alpha=r,\varphi,z}\!\!\!\nabla_{\alpha}\,^{\!\!\perp}G_{\alpha
z}(\mathbf{r},\mathbf{R},\omega)=0
\label{Gperpgaugeapp}
\end{equation}
and
\begin{equation}
\sum_{\beta,\gamma=r,\varphi,z}\!\!\!
\epsilon_{\alpha\beta\gamma}\nabla_{\beta}\;^{\!\parallel}G_{\gamma
z}(\mathbf{r},\mathbf{R},\omega)=0\,,
\label{Gpargaugeapp}
\end{equation}
where $\epsilon_{\alpha\beta\gamma}$ is the totally antisymmetric
unit tensor of rank 3.~Keeping Eqs.~(\ref{Gperpgaugeapp}) and
(\ref{Gpargaugeapp}) in mind, Eq.~(\ref{Greenperparapp}) is
rewritten to give two independent equations for the transverse and
longitudinal Green tensors of the form
\begin{eqnarray}
\left(\Delta+k^{2}\right)\,^{\!\!\perp}G_{zz}(\mathbf{r},\mathbf{R},\omega)
=-\delta_{zz}^{\perp}(\mathbf{r}-\mathbf{R})\,,
\label{Gzzperpapp}\\
k^{2}\;^{\!\parallel}G_{zz}(\mathbf{r},\mathbf{R},\omega)
=-\delta_{zz}^{\parallel}(\mathbf{r}-\mathbf{R})~~
\label{Gzzparapp}
\end{eqnarray}
with the delta-functions given by Eqs.~(\ref{deltapar}) and
(\ref{deltaper}) and $^{\!\perp(\parallel)}G_{zz}$ defined by
Eq.~(\ref{Gzzperpar}).

The transverse Green function
$^{\!\perp}G_{zz}(\mathbf{r},\mathbf{r}_{A},\omega)$ was derived
and analyzed in our previous work~\cite{Bondarev04PRB}.~Its
differential form equivalent to Eq.~(\ref{Gzzperpar}) is given by
\begin{equation}
^{\!\perp}G_{zz}(\mathbf{r},\mathbf{r}_{A},\omega)=\left(
{1\over{k^{2}}}\nabla_{z}\nabla_{z}+1\right)g(\mathbf{r},\mathbf{r}_{A},\omega)
\label{gperapp}
\end{equation}
with $g(\mathbf{r},\mathbf{r}_{A},\omega)$ being the Green
function of the scalar Helmholtz equation, satisfying the
radiation condition at infinity and boundary conditions on the CN
surface.~Indeed, substituting Eq.~(\ref{gperapp}) into
Eq.~(\ref{Gzzperpapp}) and using Eq.~(\ref{Gperpgaugeapp}), one
straightforwardly obtains
\begin{equation}
\left(\Delta+k^{2}\right)g(\mathbf{r},\mathbf{r}_{A},\omega)=
-\delta_{zz}^{\perp}(\mathbf{r}-\mathbf{r}_{A})\,,
\label{gpereqapp}
\end{equation}
the scalar Helmholtz equation with a transverse $\delta$-source.
For the longitudinal Green function
$^{\!\parallel}G_{zz}(\mathbf{r},\mathbf{r}_{A},\omega)$, one
analogously has
\begin{equation}
^{\!\parallel}G_{zz}(\mathbf{r},\mathbf{r}_{A},\omega)=
-{1\over{k^{2}}}\nabla_{z}\nabla_{z}\,g(\mathbf{r},\mathbf{r}_{A},\omega)\,,
\label{gparapp}
\end{equation}
which upon substituting into Eq.~(\ref{Gzzparapp}) yields
\begin{equation}
\nabla_{z}\nabla_{z}\,g(\mathbf{r},\mathbf{r}_{A},\omega)=
\delta_{zz}^{\parallel}(\mathbf{r}-\mathbf{r}_{A})\,.
\label{gpareqapp}
\end{equation}

The complete solution of Eq.~(\ref{gperapp}), which satisfies the
radiation condition at infinity and boundary conditions on the CN
surface, has for $r_{A}\!\ge\!r\!>\!R_{cn}$ the
form~\cite{Bondarev04PRB}
\begin{widetext}
\begin{equation}
g(\mathbf{r},\mathbf{r}_{A},\omega)=g_{0}(\mathbf{r},\mathbf{r}_{A},\omega)
-{R_{cn}\over{(2\pi)^{2}}}\sum_{p=-\infty}^{\infty}\!\!
e^{ip\varphi}\!\!\int_{C}{\beta(\omega)\,v^{2}I_{p}^{2}(vR_{cn})
K_{p}(vr_{A})K_{p}(vr)\over{1+\beta(\omega)\,v^{2}R_{cn}
I_{p}(vR_{cn})K_{p}(vR_{cn})}}\;e^{ihz}dh\,,
\label{gexplicitapp}
\end{equation}
\end{widetext}
where the coordinate system has been fixed as is shown in
Fig.~\ref{fig1}, so that $\mathbf{r}_{A}=\{r_{A},0,0\}$ and
$\mathbf{r}=\{r,\varphi,z\}$, the first item is the particular
solution of the inhomogeneous equation which is
\begin{equation}
g_{0}(\mathbf{r},\mathbf{r}_{A},\omega)={1\over{4\pi}}
{e^{ik|\mathbf{r}-\mathbf{r}_{A}|}\over{|\mathbf{r}-\mathbf{r}_{A}|}}\,,
\label{g0app}
\end{equation}
the point-source function (see, e.g.,~Ref.~\cite{Davydov}), and
the second item is the general solution of the homogeneous
equation with the boundary conditions imposed on the CN surface
and at infinity.~This latter item is represented in terms of the
modified cylindric Bessel functions $I_{p}$ and $K_{p}$ with
$v=v(h,\omega)=\sqrt{h^{2}-k^{2}}$ and $\beta(\omega)=4\pi
i\sigma_{zz}(R_{cn},\omega)/\omega$. The integration contour $C$
goes along the real axis of the complex plane and envelopes the
branch points $\pm k$ of the integrand from below and from above,
respectively. The function $g(\mathbf{r},\mathbf{r}_{A},\omega)$
for $r\!\le\!r_{A}\!<\!R_{cn}$ is obtained from
Eq.~(\ref{gexplicitapp}) by means of a simple symbol replacement
$I_{p}\!\leftrightarrow\!K_{p}$ in the numerator of the integrand.

Now, using Eqs.~(\ref{gperapp}), (\ref{gparapp}),
(\ref{Gperpgaugeapp}) and the property
$g(\mathbf{r},\mathbf{r}_{A},\omega)=g(\mathbf{r}_{A},\mathbf{r},\omega)$
which is obvious from Eqs.~(\ref{gexplicitapp}) and (\ref{g0app}),
it is not difficult to prove that
\begin{eqnarray}
^{\perp}G_{zz}^{\perp}(\mathbf{r}_{A},\mathbf{r}_{A},\omega)&=&
^{\perp}G_{zz}(\mathbf{r}_{A},\mathbf{r}_{A},\omega)\,,\label{Gzzperperapp}\\
\nonumber\\[-0.3cm]
^{\parallel}G_{zz}^{\parallel}(\mathbf{r}_{A},\mathbf{r}_{A},\omega)&=&
^{\parallel}G_{zz}(\mathbf{r}_{A},\mathbf{r}_{A},\omega)\,,\label{Gzzparparapp}\\
\nonumber\\[-0.4cm]
^{\perp}G_{zz}^{\parallel}(\mathbf{r}_{A},\mathbf{r}_{A},\omega)&=&0\,,\label{Gzzperparapp}
\end{eqnarray}
whereupon, making use of the definitions (\ref{DOSdef}),
(\ref{Gammaperpar}) and Eqs.~(\ref{G0})--(\ref{xipar}), one
arrives for $r_{A}>R_{cn}$ at the the transverse and longitudinal
local photonic DOS functions of the form
\begin{widetext}
\begin{eqnarray}
\overline{\xi}^{\,\perp}(\mathbf{r}_{A},\omega)=
{3R_{cn}\over{2\pi k^{3}}}\,\mbox{Im}\!\!\!
\sum_{p=-\infty}^{\infty}\int_{C}{dh\,\beta(\omega)\,v^{4}
I_{p}^{2}(vR_{cn})K_{p}^{2}(vr_{A})\over{1+\beta(\omega)\,v^{2}R_{cn}
I_{p}(vR_{cn})K_{p}(vR_{cn})}}\,,\label{xiperapp}\\
\overline{\xi}^{\,\parallel}(\mathbf{r}_{A},\omega)={3R_{cn}\over{2\pi
k^{3}}}\,\mbox{Im}\!\!\!\sum_{p=-\infty}^{\infty}\int_{C}{dh\,\beta(\omega)\,h^{2}v^{2}
I_{p}^{2}(vR_{cn})K_{p}^{2}(vr_{A})\over{1+\beta(\omega)\,v^{2}R_{cn}
I_{p}(vR_{cn})K_{p}(vR_{cn})}}\,.\label{xiparapp}
\end{eqnarray}\vskip1cm
\end{widetext}
Here, the sign in front of Eq.~(\ref{xiparapp}) has been changed
to be positive.~This reflects the fact that the right-hand sides
of Eqs.~(\ref{gpereqapp}) and (\ref{gpareqapp}) are of opposite
signs.~As a consequence, partial solutions of the corresponding
homogeneous equations should be taken to have opposite signs as
well.~This yields a correct (positive) sign of the longitudinal
local photonic DOS which, along with the transverse local photonic
DOS, must be a positively defined function.~For $r_{A}<R_{cn}$,
Eqs.~(\ref{xiperapp}) and (\ref{xiparapp}) should be modified by
the replacement $r_{A}\!\leftrightarrow\!R_{cn}$ in the Bessel
function arguments in the numerators of the integrands.~In
dimensionless variables (\ref{dimless}), these equations are
rewritten to give Eqs.~(\ref{ksiper}) and (\ref{ksipar}).

Finally, based upon the properties
(\ref{Gzzperperapp})--(\ref{Gzzperparapp}), one can easily prove
Eqs.~(\ref{DOS}) and (\ref{DOSparper}) for the dipole interaction
matrix elements.~The proof of Eq.~(\ref{DOS}) is
straightforward.~In view of Eq.~(\ref{gperppar}), one has
\begin{eqnarray}
\int\!d\mathbf{R}\,|\mbox{g}^{\perp(\parallel)}(\mathbf{r}_{A},\mathbf{R},\omega)|^{2}
\!\!\!&=&\!\!\!\pi\hbar\omega{16\omega_{A}^{2}d_{z}^{\,2}\over{c^{4}}}\!\int\!d\mathbf{R}\,
\mbox{Re}\,\sigma_{zz}(\mathbf{R},\omega)\nonumber\\
\times\;^{\perp(\parallel)}G_{zz}(\mathbf{r}_{A},\mathbf{R},\omega)
\mbox{\hskip-0.5cm}&&\mbox{\hskip-0.15cm}
^{\perp(\parallel)}G_{zz}(\mathbf{r}_{A},\mathbf{R},\omega)^{\!\!^{\displaystyle\ast}}.
\label{gperparapp}
\end{eqnarray}
Using further the integral relationship
\begin{eqnarray}
\mbox{Im}\,G_{\alpha\beta}(\mathbf{r},\mathbf{r}^{\prime},\omega)=
{4\pi\over{c}}\,k\mbox{\hskip1.5cm}&&\nonumber\\
\times\!\!\int\!d\mathbf{R}\,\mbox{Re}\,\sigma_{zz}(\mathbf{R},\omega)
G_{\alpha z}(\mathbf{r},\mathbf{R},\omega)G_{\beta
z}(\mathbf{r}^{\prime},\mathbf{R},\omega)^{\!\!^{\displaystyle\ast}}&&
\label{intrelation}
\end{eqnarray}
[which is nothing but a particular 2D case of a general integral
relationship proven for any 3D electromagnetic field Green tensor
in Ref.~\cite{Welsch}, with Eq.~(\ref{sigmaCN}) taken into
account], one obtains for the right-hand side of
Eq.~(\ref{gperparapp})
\[
\hbar{4\omega_{A}^{2}d_{z}^{\,2}\over{c^{2}}}\,
\mbox{Im}^{\perp(\parallel)}G^{\perp(\parallel)}_{zz}
(\mathbf{r}_{A},\mathbf{r}_{A},\omega),
\]
whereupon using Eqs.~(\ref{DOSdef}), (\ref{Gammaperpar}) and
(\ref{G0}), one arrives at Eq.~(\ref{DOS}).

The proof of Eq.~(\ref{DOSparper}) is based upon
Eq.~(\ref{DOS}).~In view of this equation, the right-hand side of
Eq.~(\ref{DOSparper}) takes the form
\begin{eqnarray}
&&\int\!d\mathbf{R}\,|\mbox{g}^{(\pm)}(\mathbf{r}_{A},\mathbf{R},\omega)|^{2}=
{(\hbar\omega_{A})^{2}\over{2\pi\omega^{2}}}\,\Gamma_{0}(\omega)\nonumber\\
&&\mbox{\hskip0.3cm}\times\left[\xi^{\perp}(\mathbf{r}_{A},\omega)+
\!\left({\omega\over{\omega_{A}}}\right)^{\!2}
\xi^{\parallel}(\mathbf{r}_{A},\omega)\right]\label{DOSparperapp}\\
&\pm&\!\!\!{\omega\over{\omega_{A}}}\,2\mbox{Re}\!
\int\!d\mathbf{R}\,\mbox{g}^{\perp}(\mathbf{r}_{A},\mathbf{R},\omega)\,
\mbox{g}^{\parallel}(\mathbf{r}_{A},\mathbf{R},\omega)^{\!\!^{\displaystyle\ast}}\!,\nonumber
\end{eqnarray}
where the last item can be further rewritten in terms of
Eqs.~(\ref{gperppar}), (\ref{Gzzperpar}) and (\ref{intrelation})
to give
\[
\hbar{8\omega\omega_{A}d_{z}^{\,2}\over{c^{2}}}\,
\mbox{Im}\,^{\perp}G^{\parallel}_{zz}(\mathbf{r}_{A},\mathbf{r}_{A},\omega)\,,
\]
which is zero, according to Eq.~(\ref{Gzzperparapp}).~Thus,
one arrives at Eq.~(\ref{DOSparper}).

\vspace{1cm}

\section{The relation of Eq.(\ref{Evw}) with the Casimir-Polder formula}\label{appC}

In deriving an 'infinitely conducting plane' result (the
Casimir-Polder formula~\cite{Casimir}) from our theory, we will
follow a~general line of the work by Marvin and
Tiogo~\cite{Marvin} who did the same within the framework of their
linear response theory.

We start with our weak-coupling-regime equation~(\ref{Evwweak1}).
First of all one has to simplify the local DOS functions
$\overline{\xi}^{\,\perp(\parallel)}(\mathbf{r}_{A},x)$ in this
equation by putting
$\overline{\sigma}_{zz}\!\rightarrow\infty$.~From
Eqs.~(\ref{ksiper}) and (\ref{ksipar}), one has then for
$r_{A}>R_{cn}$
\begin{widetext}
\begin{equation}
\left\{\!\begin{array}{c}
\overline{\xi}^{\,\perp}(\mathbf{r}_{A},x)\\
\overline{\xi}^{\,\parallel}(\mathbf{r}_{A},x)
\end{array}\!\right\}={3\over{\pi}}\;\mbox{Im}\!\int_{0}^{\infty}\!\!\!
dy\left\{\!\!\begin{array}{c}y^{2}-1-i\varepsilon\\y^{2}\end{array}
\!\!\right\}\!\sum_{p=-\infty}^{\infty}\!\!\!{K_{p}^{2}[\sqrt{y^{2}-1-i\varepsilon}\,u(r_{A})x]
I_{p}[\sqrt{y^{2}-1-i\varepsilon}\,u(R_{cn})x]\over{K_{p}[x\sqrt{y^{2}-1-i\varepsilon}\,u(R_{cn})x]}}\,,
\label{appC1}
\end{equation}
\end{widetext}
where $\varepsilon$ is an infinitesimal positive constant which is
necessary to correctly envelope the branch points of the integrand
in integrating over $y$.

The next step is taking a large radius limit in Eq.~(\ref{appC1}).
This can be done by using the relationship
\[
\lim_{\begin{array}{c}\mbox{\scriptsize$a\!\rightarrow\!\infty$}\\[-0.15cm]
\mbox{\scriptsize$b\!\rightarrow\!\infty$}\\[-0.15cm]
\mbox{\scriptsize$(a\!-\!b=const)$}\end{array}}\!\!\!\!\!\!\!
\sum_{n=-\infty}^{\infty}\!\!\!{K_{n}^{2}(a)I_{n}(b)\over{K_{n}(b)}}=K_{0}[2(a-b)]
\]
proved in Ref.~\cite{Marvin}.~One has
\begin{eqnarray}
\left\{\!\begin{array}{c}
\overline{\xi}^{\,\perp}(\mathbf{r}_{A},x)\\
\overline{\xi}^{\,\parallel}(\mathbf{r}_{A},x)\end{array}\!\right\}
={3\over{\pi}}\;\mbox{Im}\!\int_{0}^{\infty}\!\!\!dy\left\{\!\!
\begin{array}{c}y^{2}-1-i\varepsilon\\y^{2}\end{array}\!\!\right\}
\nonumber\\[0.1cm]
\times\,K_{0}[2\mu x\sqrt{y^{2}-1-i\varepsilon}\;]
\,,\mbox{\hskip0.5cm} \label{appC2}
\end{eqnarray}
where $\mu\!=\!2\gamma_{0}l/\hbar c$ with $l\!=\!r_{A}-R_{cn}$
being the atom-surface distance.~The integration variable $y$ in
this equation contains implicit frequency dependence.~Indeed,
comparing dimensionless equations (\ref{ksiper}), (\ref{ksipar})
with (\ref{xiperapp}), (\ref{xiparapp}), one can see that
$y=h/k=h(c/\omega)=h(c\hbar/2\gamma_{0}x)$. For the following it
makes sense to extract this frequency dependence by the
substitution $y\!=\!\tau/x$ with dimensionless
$\tau=h(c\hbar/2\gamma_{0})$.~The result takes the form
\begin{equation}
\overline{\xi}^{\,\perp(\parallel)}(\mathbf{r}_{A},x)=
{3\over{2i\pi x}}\left\{f^{\perp(\parallel)}(\mathbf{r}_{A},x)-
\!f^{\perp(\parallel)}(\mathbf{r}_{A},x)^{\!\!^{\displaystyle\ast}}\right\}
\label{appC3}
\end{equation}
with
\begin{eqnarray}
\left\{\!\begin{array}{c}
f^{\perp}(\mathbf{r}_{A},x)\\
f^{\parallel}(\mathbf{r}_{A},x)\end{array}\!\right\}
=\int_{0}^{\infty}\!\!\!d\tau\left\{\!\!
\begin{array}{c}(\tau/x)^{2}-1-i\varepsilon\\
(\tau/x)^{2}\end{array}\!\!\right\}
\nonumber\\[0.25cm]
\times\,K_{0}[2\mu x\sqrt{(\tau/x)^{2}-
1-i\varepsilon}\;]\,.\mbox{\hskip0.5cm} \label{appC4}
\end{eqnarray}

Substituting Eq.~(\ref{appC3}) into Eq.~(\ref{Evwweak1}), one has
\begin{eqnarray}
\varepsilon_{vw}(\mathbf{r}_{A})={3x_{A}\over{4i\pi^{2}}}\mbox{\hskip2.5cm}\nonumber\\
\times\!\left\{\int_{0}^{\infty}\!\!\!{dx\,\tilde{\Gamma}_{0}(x)\over{(x_{A}\!+x)x^{2}}}\!
\left[f^{\perp}(\mathbf{r}_{A},x)\!-\!{x\over{x_{A}}}f^{\parallel}(\mathbf{r}_{A},x)\right]
\right.\mbox{\hskip0.5cm}\label{appC5}\\
-\!\left.\int_{0}^{-\infty}\!\!\!{dx\,\tilde{\Gamma}_{0}(x)\over{(x_{A}\!-x)x^{2}}}\!
\left[f^{\perp}(\mathbf{r}_{A},-x)^{\!\!^{\displaystyle\ast}}\!+\!
{x\over{x_{A}}}f^{\parallel}(\mathbf{r}_{A},-x)^{\!\!^{\displaystyle\ast}}\right]
\!\right\}.\nonumber
\end{eqnarray}
Here, in the second item, the change of the integration variable
from $x$ to $-x$ has been made and it has been taken into account
that $\tilde{\Gamma}_{0}(-x)\!=\!-\tilde{\Gamma}_{0}(x)$.~Noting
further that both integrands in Eq.~(\ref{appC5}) do not have
poles in the upper complex half plane, one may rotate the
integration path of the first and the second integral by $\pi/2$
and by $-\pi/2$, respectively.~Then, both integrals become the
integrals over the positive imaginary axis of the complex plane
and Eq.~(\ref{appC5}) takes the form
\begin{eqnarray}
\varepsilon_{vw}(\mathbf{r}_{A})={3x_{A}\over{4i\pi^{2}}}\mbox{\hskip2.5cm}\nonumber\\
\times\!\left\{\int_{0}^{i\infty}\!\!\!{dz\,\tilde{\Gamma}_{0}(z)\over{(x_{A}\!+z)z^{2}}}\!
\left[f^{\perp}(\mathbf{r}_{A},z)\!-\!{z\over{x_{A}}}f^{\parallel}(\mathbf{r}_{A},z)\right]
\right.\mbox{\hskip0.5cm}\label{appC6}\\
-\!\left.\int_{0}^{i\infty}\!\!\!{dz\,\tilde{\Gamma}_{0}(z)\over{(x_{A}\!-z)z^{2}}}\!
\left[f^{\perp}(\mathbf{r}_{A},-z)^{\!\!^{\displaystyle\ast}}\!+\!
{z\over{x_{A}}}f^{\parallel}(\mathbf{r}_{A},-z)^{\!\!^{\displaystyle\ast}}\right]
\!\right\}.\nonumber
\end{eqnarray}
Making further the change of the integration variable from $z$ to
$iu$ with real non-negative $u$, one arrives at
\begin{eqnarray}
\varepsilon_{vw}(\mathbf{r}_{A})={3x_{A}\over{4\pi^{2}}}\mbox{\hskip2.5cm}\nonumber\\
\times\!\left\{-\!\int_{0}^{\infty}\!\!\!{du\,\tilde{\Gamma}_{0}(iu)\over{(x_{A}\!+iu)u^{2}}}\!
\left[f^{\perp}(\mathbf{r}_{A},iu)\!-\!{iu\over{x_{A}}}f^{\parallel}(\mathbf{r}_{A},iu)\right]
\right.\mbox{\hskip0.5cm}\label{appC7}\\
+\!\left.\int_{0}^{\infty}\!\!\!{du\,\tilde{\Gamma}_{0}(iu)\over{(x_{A}\!-iu)u^{2}}}\!
\left[f^{\perp}(\mathbf{r}_{A},-iu)^{\!\!^{\displaystyle\ast}}\!+\!
{iu\over{x_{A}}}f^{\parallel}(\mathbf{r}_{A},-iu)^{\!\!^{\displaystyle\ast}}\right]
\!\right\}.\nonumber
\end{eqnarray}
Here, the functions $f^{\perp(\parallel)}(\mathbf{r}_{A},iu)$ have
the following explicit form
\begin{eqnarray}
\left\{\!\!\begin{array}{c}
f^{\perp}(\mathbf{r}_{A},iu)\\
f^{\parallel}(\mathbf{r}_{A},iu)\end{array}\!\!\right\}\!
=-u\!\!\int_{1}^{\infty}\!\!\!\!\!d\eta\left\{\!\!
\begin{array}{c}\eta^{3}\!/\sqrt{\eta^{2}-1}\\
\eta\sqrt{\eta^{2}-1}\end{array}\!\right\}\!K_{0}(2\mu u\eta)
\label{appC8}
\end{eqnarray}
obtained from Eq.~(\ref{appC4}) after the change of the
integration variable by the subsequent substitutions $x=iu$ and
$\eta^{2}\!=\!\tau^{2}/u^{2}+1$.~The functions
$f^{\perp(\parallel)}(-iu)^{\ast}\! =\!f^{\perp(\parallel)}(iu)$
as a~consequence~of a general property
$K_{\nu}(z)^{\ast}\!=\!K_{\nu}(z^{\ast})$~\cite{Abramovitz}.
Substituting Eq.~(\ref{appC8}) into Eq.~(\ref{appC7}) and making
allowance for the fact that
$\tilde{\Gamma}_{0}(iu)\!=\!-i\tilde{\Gamma}_{0}(u)$, one further
has
\begin{eqnarray}
\varepsilon_{vw}(\mathbf{r}_{A})=-{3x_{A}\over{2\pi^{2}}}\!
\int_{0}^{\infty}\!{du\,\tilde{\Gamma}_{0}(u)\over{x_{A}^{2}\!+u^{2}}}\mbox{\hskip0.75cm}\nonumber\\
\times\!\int_{1}^{\infty}\!\!\!d\eta\,\eta\!\left(\!\sqrt{\eta^{2}-1}
+{\eta^{2}\over{\sqrt{\eta^{2}-1}}}\right)\!K_{0}(2\mu u\eta)\,,
\label{appC9}
\end{eqnarray}
which upon the substitution $\eta^{2}=\rho$ becomes
\begin{eqnarray}
\varepsilon_{vw}(\mathbf{r}_{A})=-{3x_{A}\over{4\pi^{2}}}\!
\int_{0}^{\infty}\!{du\,\tilde{\Gamma}_{0}(u)\over{x_{A}^{2}\!+u^{2}}}\mbox{\hskip0.75cm}\nonumber\\
\times\!\int_{1}^{\infty}\!\!\!d\rho\left(\!\sqrt{\rho-1}
+{\rho\over{\sqrt{\rho-1}}}\right)\!K_{0}(2\mu u\sqrt{\rho}\,)\,.
\label{appC10}
\end{eqnarray}

In Eq.~(\ref{appC10}), the first integral over $\rho$ is taken
exactly by means of the relationship~\cite{Ryzhik}
\[
\int_{1}^{\infty}\!\!\!\!\!d\rho\sqrt{\rho-1}\,K_{0}(a\sqrt{\rho}\,)\!=\!
\sqrt{2\pi\over{a^{3}}}\,K_{3/2}(a)
\longrightarrow_{^{^{^{^{^{^{^{\mbox{\hskip-0.75cm}}}}}}}}a\rightarrow\infty}
{\pi\over{a^{3}}}(1+a)e^{-a}
\]
and the second one -- by making a large argument series expansion
of the integrand and its subsequent termwise integration.~This,
upon returning back to dimensional variables by means of
Eqs.~(\ref{dimless}) and using the definition (see,
e.g.,~\cite{Agarwal})
\[
\alpha_{zz}(iu)={2\over{\hbar}}\,{d_{z}^{\,2}\,\omega_{A}\over{\omega_{A}^{2}-(iu)^{2}}}
\]
for the polarizability tensor $zz$-component of the two-level
system, finally yields
\begin{eqnarray*}
E_{vw}(\mathbf{r}_{A})&\approx&-{\hbar\over{8\pi
l^{3}}}\int_{0}^{\infty}\!\!\!du\,\alpha_{zz}(iu)\\
&\times&\left[1+2{u\over{c}}\,l+2\left({u\over{c}}\right)^{2}l^{2}\right]e^{-2(u/c)l}.
\end{eqnarray*}
This is the half of the well-known Casimir-Polder result for the
vdW energy of an atom near an infinitely conducting plane (see
Refs.~\cite{Casimir,Marvin}), in agreement with the fact that our
model of the atom-electromagnetic-field interactions in the
presence of a nanotube only takes the longitudinal (along the
nanotube axis, or, equivalently, parallel to the plane in the
large nanotube radius limit) atomic dipole orientation into
account.~Another half of the Casimir-Polder vdW energy would come
from the transverse (perpendicular to the plane) atomic dipole
orientation which we have neglected.

\end{document}